\documentclass{aa}  
\usepackage{graphicx}
\usepackage{txfonts}
\usepackage{hyperref}
\hypersetup{
     colorlinks   = true,
     citecolor    = blue
}

\usepackage{siunitx}
\usepackage{gensymb}
\usepackage{caption,subcaption}
\usepackage{subcaption}
\usepackage{float}
\usepackage{pgffor}
\usepackage{dblfloatfix}    
\def \Ha{H$\alpha$}
\def \agc{AGC 226178}
\def \msun{$\,M_{\odot}\,$}
\def \mstar{$\,M_{\star}\,$}
\def \chisq{$\chi^2$}
\def \magperarcsec{mag arcsec$^{-2}$}
\def \udg{NGVS 3543}

\begin{document} 

\title{A Virgo Environmental Survey Tracing Ionised Gas Emission (VESTIGE).X. Formation of a red ultra-diffuse galaxy and an almost dark galaxy during a ram-pressure stripping event\thanks{Based on observations obtained with MegaPrime/MegaCam, a joint project of CFHT and CEA/DAPNIA, at the Canada-French-Hawaii Telescope (CFHT) which is operated by the National Research Council (NRC) of Canada, the Institut National des Sciences de l'Univers of the Centre National de la Recherche Scientifique (CNRS) of France and the University of Hawaii.}$^{,}$\thanks{The images (FITS files) are available at the CDS via anonymous ftp to cdsarc.u-strasbg.fr (130.79.128.5) or via \href{http://cdsweb.u-strasbg.fr/cgi-bin/qcat?J/A+A/} {http://cdsweb.u-strasbg.fr/cgi-bin/qcat?J/A+A/}}}

    \titlerunning{Formation of a red UDG and an "almost dark" galaxy during a ram-pressure stripping event}

   \author{
            Junais\inst{1}, 
            S. Boissier\inst{1},
            A. Boselli\inst{1},
            M. Boquien\inst{2},            
            A. Longobardi\inst{1},
            Y. Roehlly\inst{1},
            P. Amram\inst{1},
            M. Fossati\inst{3},
            J.-C. Cuillandre\inst{4}, 
            S. Gwyn\inst{5},
            L. Ferrarese\inst{5},
            P. C{\^o}t{\'e}\inst{5},
            J. Roediger\inst{5},
            S. Lim\inst{5,6},
            E.W. Peng\inst{7,8},
            G. Hensler\inst{9},
            G. Trinchieri\inst{10},
            J. Koda\inst{11},
            N. Prantzos\inst{12}
            }
    \authorrunning{Junais et al.}
  \institute{Aix Marseille Univ, CNRS, CNES, LAM, Marseille, France\\   
             \email{junais.madathodika@lam.fr, samuel.boissier@lam.fr, alessandro.boselli@lam.fr}
             \and
            Centro de Astronom\'a (CITEVA), Universidad de Antofagasta, Avenida Angamos 601, Antofagasta, Chile 
           \and
            Dipartimento di Fisica G. Occhialini, Universit\`a degli Studi di Milano-Bicocca, Piazza della Scienza 3, 20126 Milano, Italy 
            \and
            AIM, CEA, CNRS, Universit\'{e} Paris-Saclay, Universit\'{e}  Paris Diderot, F-91191 Gif-sur-Yvette, France 
            \and
            NRC Herzberg Astronomy and Astrophysics, 5071 West Saanich Road, Victoria, BC, V9E 2E7, Canada 
            \and
            University of Tampa, 401 West Kennedy Boulevard, Tampa, FL 33606, USA 
            \and
            Department of Astronomy, Peking University, Beijing 100871, PR China 
            \and
            Kavli Institute of Astronomy and Astrophysics, Peking University, Beijing 100871, PR China  
           \and
           Department of Astrophysics, University of Vienna, T\"urkenschanzstrasse 17, 1180 Vienna, Austria 
           \and
           INAF - Osservatorio Astronomico di Brera, via Brera 28, 20159 Milano, Italy 
            \and
            Department of Physics and Astronomy, Stony Brook University, Stony Brook, NY 11794-3800, USA 
            \and
            Institut d’Astrophysique de Paris, UMR7095 CNRS, Sorbonne Universit\'{e}, 98bis Bd. Arago, 75104 Paris, France 
             }

   \date{Received 21 December 2020 / Accepted 17 March 2021}

 
  \abstract
  {The evolution of galaxies depends on their  interaction with the surrounding environment. 
  Ultra-diffuse galaxies (UDGs) have been found in large numbers in clusters.  We detected a few star-forming blobs in the VESTIGE survey, located at $\sim$5 kpc from a UDG, namely \udg{}, in association with an HI gas cloud \agc{}, suggesting a recent interaction between this low-surface-brightness system and the surrounding cluster environment.
}
  {We use a complete set of multi-frequency data including deep optical, UV, and narrow-band \Ha{} imaging and HI data to understand the formation process that gave birth to this peculiar system.}
  {For this purpose, we measured (i) the multi-wavelength radial surface brightness profiles of \udg{} and compared them to the predictions of spectro-photometric models of galaxy evolution in rich clusters; and (ii) the aperture photometry of the blue regions in the vicinity of \udg{} in order to determine their age and stellar mass.
}
  {Comparisons of the observations with evolutionary models indicate that  \udg{} has undergone a ram-pressure stripping (RPS) that peaked $\sim$100 Myr ago, transforming a blue gas-rich UDG into a red gas-poor UDG. Star formation has taken place in the ram pressure stripped gas, the mass of which is $\sim$10$^{8}$\msun, forming star complexes with a typical age of $\sim$20 Myr and a stellar mass of $\sim$10$^4$\msun.}
%
  {These results suggest that we are observing for the first time the ongoing transformation of a gas-rich UDG into a red and quiescent UDG under the effect of a ram-pressure stripping event. The same process could explain the lack of star-forming UDGs in rich environments observed in several nearby clusters.}
   \maketitle
%

\section{Introduction}

\begin{figure*}
\centering
\includegraphics[width=\hsize]{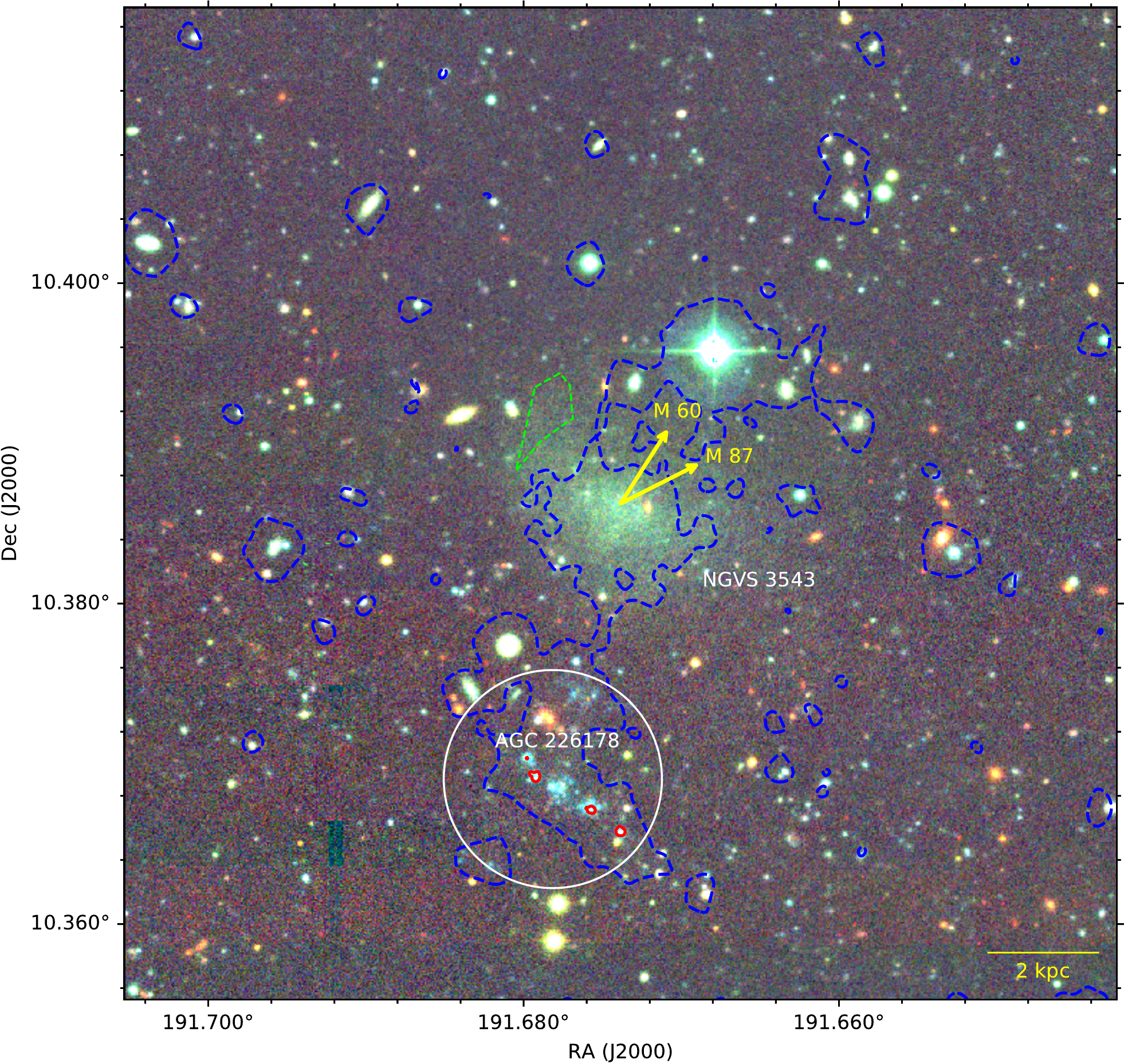}
  \caption{NGVS \textit{u, g, i}-colour composite image of the UDG galaxy \udg{}.
  The yellow arrows indicate the direction towards the Virgo cluster elliptical galaxies M87 and M60 at a distance of 1.26 Mpc and 0.39 Mpc, respectively. Blue dashed contours indicate the GALEX NUV detection at a surface brightness level of 27 \magperarcsec{} and red contours indicate the \Ha{} detection in VESTIGE at the level of $1.6\times10^{-17}$ erg s$^{-1}$ cm$^{-2}$ arcsec$^{-2}$ (3$\sigma$). The white circle marks the position of the VLA HI detection of the source \agc{} from \citet{cannon2015} with a beam size of 49$\arcsec$. The green dashed region along the northeast of \udg{} shows the possible tidal feature discussed in Sect. \ref{global_scenario}.}
     \label{color_image}
\end{figure*}

The Virgo cluster is one of the richest clusters of galaxies in the nearby Universe, making it a prime candidate for deep, blind surveys at all wavelengths.
Owing to the depth of surveys like the NGVS (Next Generation Virgo cluster Survey; \citealt{ferrarese12}), VESTIGE (Virgo Environmental Survey Tracing Ionised Gas Emission; \citealt{boselli18}), and GUViCS (GALEX Ultraviolet Virgo Cluster Survey;  \citealt{boselli11}), we can now study very low-surface-brightness objects in great detail at unprecedented depths. Found in large numbers in clusters, ultra-diffuse galaxies (UDGs) are a class of galaxy that has attracted a lot of attention in recent years \citep{van_dokkum15,koda15,mihos15,vanderburg16,venhola17}. %
Although UDGs are a subset of low-surface-brightness galaxies that have been studied for decades \citep{sandage1984,caldwell1987,impey1988,conselice2003,yagi16,conselice18}, a vast number of them were found recently with deep surveys.
UDGs are often defined as galaxies with an effective radius ($R_{e}$) > 1.5 kpc and central disk surface brightness ($\mu_{0,g}$) > 24 \magperarcsec{} \citep{van_dokkum15,koda15}. 
\citet{sungsoon2020} recently defined UDGs on a more physical basis as outliers from galaxy scaling relations in the Virgo cluster. In an ongoing analysis of a large sample of low-surface-brightness galaxies (LSBs), UDGs as defined by \citet{van_dokkum15,koda15}, and UDGs as defined by \citet{sungsoon2020} in the Virgo cluster (Junais et al., in preparation), %
we noticed blue knots and diffuse emission  within a few kiloparsecs of one of our targets, the UDG NGVSJ12:46:41.73+10:23:10.4, which hereafter we refer to as \udg{} (based on the position of this galaxy in the NGVS catalog), as well as \Ha{} emission in the narrow-band image taken during the VESTIGE survey (Fig. \ref{color_image}).

Most of this emission is concentrated in blue knots close to the position of \agc{}, an HI cloud detected during ALFALFA, an HI blind survey 
also covering the Virgo cluster \citep{Giovanelli05,haynes2011}. This HI cloud, without any evident optical counterpart\footnote{\citet{cannon2015} noticed a possible UV counterpart in the GALEX images.}, was identified as an "almost dark galaxy" by \citet{cannon2015}, who made deeper targeted observations of this source with the VLA. "Dark galaxies" (galaxies with gas but no stars) were looked upon as a possible solution to the large number of small galaxies predicted by the $\Lambda$CDM cosmology \citep{verde2002}. Stars were eventually found in most candidates at low redshift, and the interest turned to "almost dark" or "near dark" objects (gas-rich objects without any clear definition; \citealt{cannon2015,janowiecki2015}). The origin of \agc{} and other almost dark galaxies has been discussed without obtaining definitive answers \citep{cannon2015,janowiecki2015,leisman2017,brunker2019}. Among the propositions for their origin are suggestions that dark galaxies are:  disks of high angular momentum (spin) that are stable against star formation \citep{jimenez2020,leisman2017}; galaxies with low star formation efficiency \citep{janowiecki2015}; galaxies that are gas stripped by or falling onto a companion galaxy \citep{sorgho2020}; or tidal debris, as in the cases of  VIRGO-HI \citep{duc2008,boselli18_vestige3} and SECCO 1 \citep{beccari2017} for example. \citet{cannon2015} classify their sample of almost dark candidates as either tidal debris or dwarf galaxies (as for \agc{}). 
In our new NGVS imaging, the optical counterpart to the elongated UV emission is resolved into very bright blue knots, several of them with detection of \Ha{} emission. More knots and diffuse emission are seen to the south of the galaxy, with a morphology similar to so-called fireball galaxies found in clusters \citep{yoshida2008}. Star-forming regions formed within the tails of ram-pressure-stripped galaxies were first discovered by \citet{gavazzi2001} in A1367. These peculiar objects are now quite commonly observed in nearby clusters provided that deep observations sensitive to the ionized gas emission are available \citep{sun2007,cortese2007,yoshida2008,yagi10,fossati2016,Poggianti2019,Gullieuszik+2020}. They have also been observed in gas-rich low-surface-brightness systems within the Virgo cluster (VCC1217, \citealt{hester2010,fumagalli2011,jachym2013,kenney2014}; IC 3476, \citealt{boselli20_vestigeIX}), but so far have not been found to be associated with dwarf quiescent galaxies.

To our knowledge, it is the first time that fireball-style knots
have been seen in connection with a UDG, making \udg{} an important object with which to distinguish some processes that have been suggested for the formation of UDGs. Indeed, many propositions have been made in recent years concerning the formation of UDGs, including the potential role of halo angular momentum, feedback, tidal interactions, ram-pressure stripping (RPS), and collisions \citep[e.g.,][]{amorisco2016,burkert17,martin2019,dicintio2019,tremmel2020,silk19}.

In this paper, we therefore analyze the full system including the UDG galaxy \udg{} and the associated blue knots in its vicinity. 
In Sect. \ref{DATA_AND_MEASUREMENTS}, we discuss details of the multi-wavelength data we use, as well as measurements performed on the images. In Sect. \ref{analysis} we present the results obtained during this study, and in Sect. 4 we provide a detailed discussion on the implications of these results along with a comparison of existing data and models. We conclude in Sect. \ref{conclusion}.

Consistently with other VESTIGE and NGVS studies, we assume the Virgo cluster to be at a distance of 16.5 Mpc \citep{gavazzi1999,mei2007,blakeslee2009}, with a projected angular scale of 80 pc arcsec$^{-1}$.
%
\section{Data and measurements}\label{DATA_AND_MEASUREMENTS}

\subsection{Data}
Our work is based on the analysis of images obtained by multi-wavelength surveys of the Virgo Cluster: the 
NGVS in the optical \citep{ferrarese12}, 
GUViCS (\citealt{boselli11}) in the UV, and 
VESTIGE (\citealt{boselli18}) for \Ha{} narrow-band and $r$-band imaging.
These surveys provide comprehensive imaging of the Virgo cluster in optical (\textit{u, g, r, i, z}, \Ha{}) and UV (far-ultraviolet (FUV) and near-ultraviolet (NUV)) bands.

VESTIGE is a blind \Ha{} narrow-band\footnote{The VESTIGE narrow-band \Ha{} filter includes the H$\alpha$ line and the two nearby [NII] emission lines at $\lambda$6548 and 6583\AA. Hereafter we refer to the \Ha{}+[NII] contribution simply as \Ha{} unless otherwise stated.} imaging survey of the Virgo cluster carried out with MegaCam at the Canada-France-Hawaii Telescope (CFHT) and was designed to cover the whole Virgo cluster up to its virial radius (104 deg$^2$). The depth and extremely high image quality of the survey makes it perfectly suitable for studying the effects of the environment on the star formation process in perturbed galaxies down to scales of $\sim$100 pc. The line sensitivity limit of the survey is $f(\text{\Ha{}})\sim 4\times10^{-17}$ erg s$^{-1}$ cm$^{-2}$ (5$\sigma$ detection limit) for point sources and \textit{$\Sigma$(\Ha{})} $\sim$2$\times10^{-18}$ erg s$^{-1}$ cm$^{-2}$ arcsec$^{-2}$ ($1\sigma$ detection limit at 3$\arcsec$ resolution) for extended sources. The contribution of the stellar continuum emission in the narrow-band \Ha{} filter is determined and removed using a combination of the $r$- and $g$-band images, as described in \citet{boselli19}. The narrow-band \Ha{} filter is optimal to detect the line emission of galaxies at the redshift of the Virgo cluster with a typical recessional velocity of $-500\leq cz \leq 3000$ km s$^{-1}$.
In the context of UDGs, VESTIGE provides information on recent star formation, but can also confirm the redshift of sources in cases of detection, whereas spectroscopy is challenging for these diffuse objects.

%
\begin{table}
\centering                          
\caption{Properties of the imaging data used in this work.}             
\begin{tabular}{c c c c}        
\hline\hline                 
Survey & Filter & FWHM & Exposure time (s)\\    
\hline                        
   NGVS & \textit{u} & 0.88$\arcsec$ & 6402 \\      
   NGVS & \textit{g} & 0.80$\arcsec$ & 3170 \\
   VESTIGE & \textit{r} & 0.65$\arcsec$ & 480 \\
   NGVS & \textit{i} & 0.54$\arcsec$ & 2055 \\
   NGVS & \textit{z} & 0.75$\arcsec$ & 3850 \\
   VESTIGE & \Ha{} & 0.64$\arcsec$ & 6600 \\
   GUViCS & NUV & $\sim$5$\arcsec$    & 3346 \\ 
   GUViCS & FUV & $\sim$5$\arcsec$    & 1632 \\ 
\hline                                   
\end{tabular}
\label{exposure_psf}      
\end{table}
%

\subsection{Radial profiles of \udg{}}\label{profile_measurements_of_udg}
\begin{figure}
\centering
\includegraphics[width=0.97\linewidth]{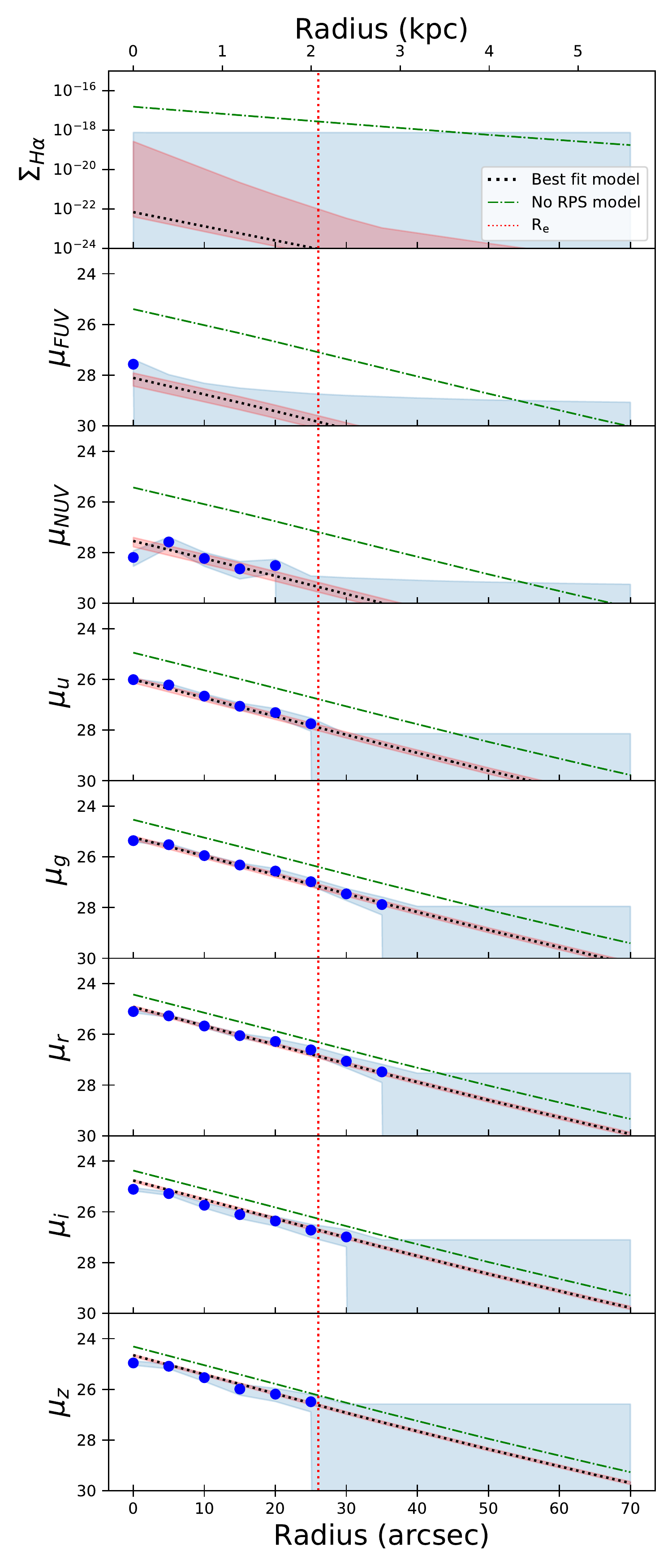}
  \caption{Radial surface-brightness profiles of \udg{} measured in eight bands, shown as blue filled dots. The surface brightness units are in \magperarcsec{} for all the bands except for \Ha{} which is in erg s$^{-1}$ cm$^{-2}$ arcsec$^{-2}$. The blue shaded area marks the 1$\sigma$ error (for data points) and upper-limits (3$\sigma$). The black dotted line indicates the best-fit model described in Sect. \ref{analysis_rps} for a ram-pressure stripped galaxy ($V = 42^{+8}_{-4}$ km s$^{-1}$, $\lambda = 0.14^{+0.02}_{-0.01}$ and $t_{rps} = 13.4\pm0.1$ Gyr). The green dot-dashed line shows the same model for an unperturbed system. The red shaded area shows the range of models allowed for the same spin and velocity, but allowing variation in the RPS efficiency and FWHM parameters as discussed in Sect. \ref{analysis_rps}. The vertical red dotted line gives our measured \textit{g}-band effective radius of the galaxy.} 
  \label{profiles_and_model}
\end{figure}

We gathered all available images of \udg{} in the optical and in the UV bands (\textit{u, g, r, i, z}, \Ha{}, FUV and NUV). We used the {\tt Montage} tool \citep{montage} to co-add all the exposures of the galaxy field in each band, projecting the new images on the pixel scale of the original NGVS and VESTIGE images (with pixels of 0.187\arcsec). 
The NGVS provides a mask for artifacts, foreground stars, stellar halos, background galaxies, and globular clusters in the field of our galaxy which was produced using multiple Sextractor runs \citep{sextractor} followed by a THELI \textit{automask} procedure \citep{theli1,theli2}. The detailed procedure followed for the NGVS mask creation is presented in \citet{ngvs_mask_procedures}. We manually edited the NGVS mask to remove residual artefacts and faint stars. Our images were then interpolated over the masked regions using the IRAF \textit{fixpix} procedure.
For the NGVS and VESTIGE images, a convolution of the above data with a Gaussian kernel was also done to match their initial resolution to that of GALEX (which we assumed to be $\mathrm{FWHM} = 5\arcsec$; see Table \ref{exposure_psf}). These images were then used to measure the radial surface brightness profiles of \udg{} shown in Fig. \ref{profiles_and_model} using the \textit{Ellipse} task in {\tt Photutils} python package \citep{photutils}, adopting the geometrical parameters for the galaxy taken from the NGVS catalog (see Table \ref{galaxy_properties}).
We also adopt the Galactic reddening from the same catalog, $E(B-V) = 0.02489$ \citep{Schlegel1998}, and correct
for Galactic extinction adopting the \citet{cardelli89} extinction curve.
We assumed that there was no internal extinction, as it is generally found in low-surface-brightness quiescent galaxies \citep{hinz,rahman}.
Error-bars in the surface brightness profile were computed by combining a pixel-scale and large-scale deviation in the sky, following the procedures given in \citet{GildePaz2005}. Profiles were measured up to three times the effective radius provided by the NGVS catalog (see Table \ref{galaxy_properties}).

The surface brightness profiles shown in Fig. \ref{profiles_and_model} %
are very close to exponential in the \textit{u}, \textit{g}, \textit{r}, \textit{i}, \textit{z} and NUV bands.
We only obtained an upper limit in the \Ha{} narrow band, and a central detection in FUV, suggesting that star formation has been low throughout the last 100 Myr. We measured the central surface brightness and effective radius of this galaxy from our profiles in the \textit{g}-band with an exponential fit in order to obtain $\mu_{0,g}=25.29$ \magperarcsec{} and $R_{e,g} = 26\farcs05$ (2.08 kpc).
\label{secmeasuredeffectiveradius}
These values are close to the NGVS ones given in Table \ref{galaxy_properties} (although the profiles were measured in slightly different ways, with a Galfit S\'ersic fit in the case of NGVS and an exponential fit in our case). This confirms that \udg{} falls under the classical definition of the UDG regime \citep{van_dokkum15,koda15}. \udg{} is not included in the definition  by \citet{sungsoon2020}, where UDGs are defined as $2.5\sigma$ outliers in scaling relationships (see Fig. 1 of \citealt{sungsoon2020}). However, we verified that \udg{} lies very close (at $2.2\sigma$) to the separation curve in these relations.

%

\begin{table}
\centering                          
\caption[]{\label{galaxy_properties}Properties of the galaxy \udg{} taken from the NGVS catalog.}
\begin{tabular}{lc}
\hline \hline
Property &
Value \\
\\ \hline
R.A. (J2000)   & 12$^\text{h}$ 46$^\text{m}$ 41.73$^\text{s}$ \\
Dec. (J2000)   & +10$^\circ$ 23$\arcmin$ 10.4$\arcsec$ \\
Distance (Mpc)        & 16.5     \\
D$_{\mathrm{M87}}$ (Mpc) & 1.26 \\
D$_{\mathrm{M60}}$ (Mpc) & 0.39 \\
Inclination angle          & $30.1^\circ  \pm 0.4^\circ$ \\
Position angle (PA)  &      $61.7^\circ  \pm 1.1^\circ$ \\
\textit{g} (mag)    &  $17.495 \pm 0.006$ \\
$\mu_{0,g}$ (\magperarcsec{}) & 25.05 \\
$R_{e,g}$ (kpc) & $1.79 \pm 0.06$ \\
\hline
\noalign{\smallskip}
\end{tabular}
\label{galaxy_properties}      
\end{table}

\subsection{Selection of blue regions}

To understand the nature of the young stellar systems associated to \udg{} and \agc{} seen in Fig. \ref{color_image}, we need first to identify them and then to characterize their spectro-photometric properties. For this purpose, we followed two different selection criteria.
The first one is based on the NGVS \textit{u}-band image, which has the advantage of having an excellent angular resolution (0.88\arcsec, corresponding to 70 pc); the second one is based on the GALEX NUV image, which despite its poorer angular resolution (5\arcsec, corresponding to 400 pc) is more sensitive to the youngest stellar population and is thus perfectly suited to identifying newly formed objects (e.g., \citealt{boselli18_vestige3}, NGC 4254).

\subsubsection{$u$-band selection}\label{u_band_selection}

We first proceeded with the identification of peaks in the NGVS $u$-band image. We used the {\tt Photutils} \textit{find\_peaks} package to identify all the peaks in $u$-band image above $5\sigma$ of the sky. The identified peaks were used as an initial set of regions, for which we performed aperture photometry within circular regions of $3\arcsec$ diameter.
The size of the aperture was optimally chosen at the same time to include the entire flux of each individual region and to resolve them from nearby companions.
The photometry was performed with the {\tt Photutils} \textit{aperture\_photometry} package in \textit{u}, \textit{g}, \textit{r}, \textit{i}, \textit{z,} and \Ha{} bands, and corrected for Milky Way foreground Galactic extinction (as described in Sect. \ref{profile_measurements_of_udg}).

\begin{figure}
\centering
\includegraphics[width=\hsize]{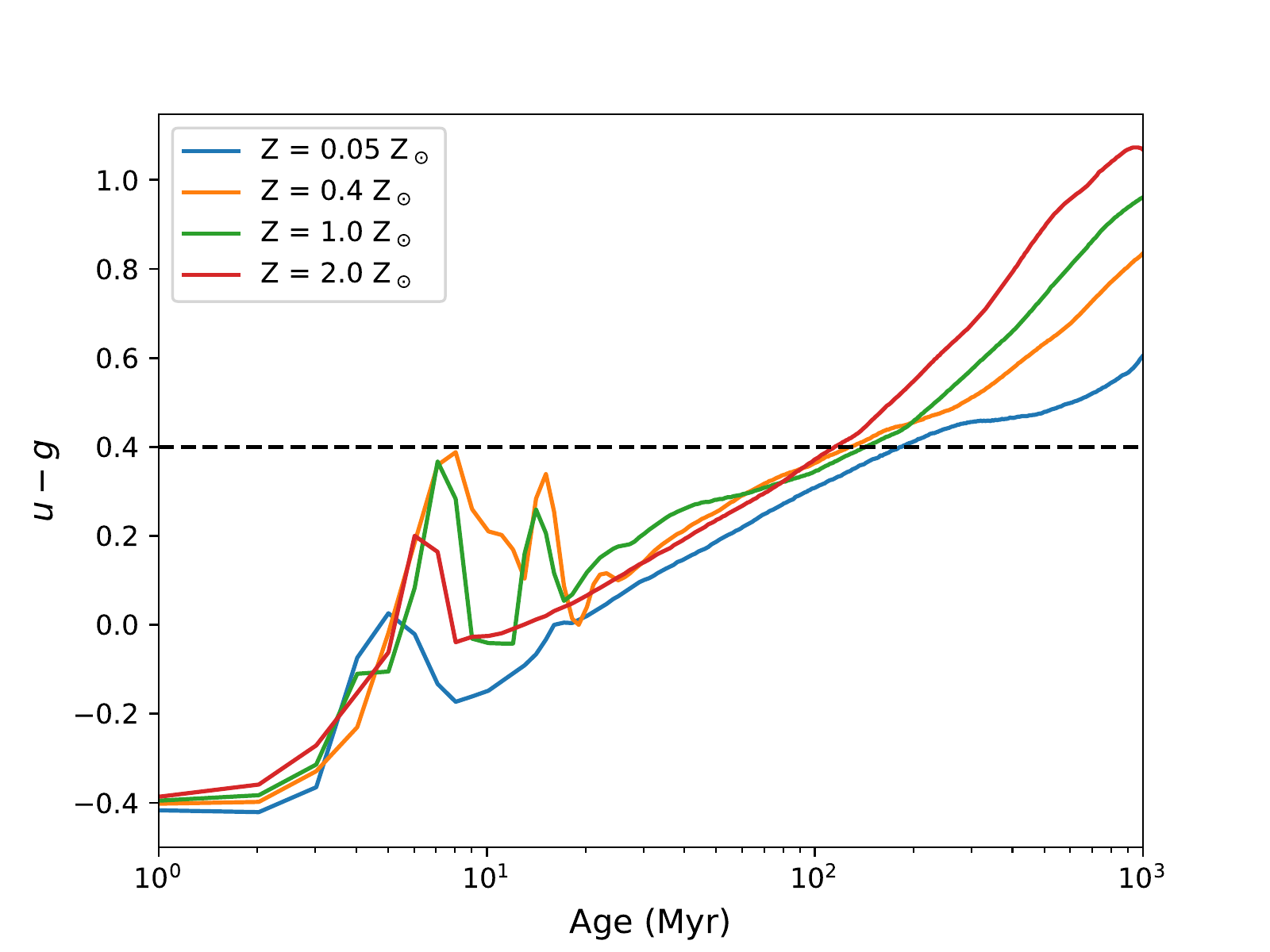}
  \caption{$u-g$ color evolution of a single-burst population derived using Starburst99 models \citep{starburst99} for different metallicities. The black dashed line marks the $u-g$ color limit (corresponding to an approximate age of less than 100 Myr) we adopted for the selection of young regions discussed in Sect. \ref{u_band_selection}.}
     \label{u-g_color_evolution_starburst99}
\end{figure}

\begin{figure}
\centering
\includegraphics[width=\hsize]{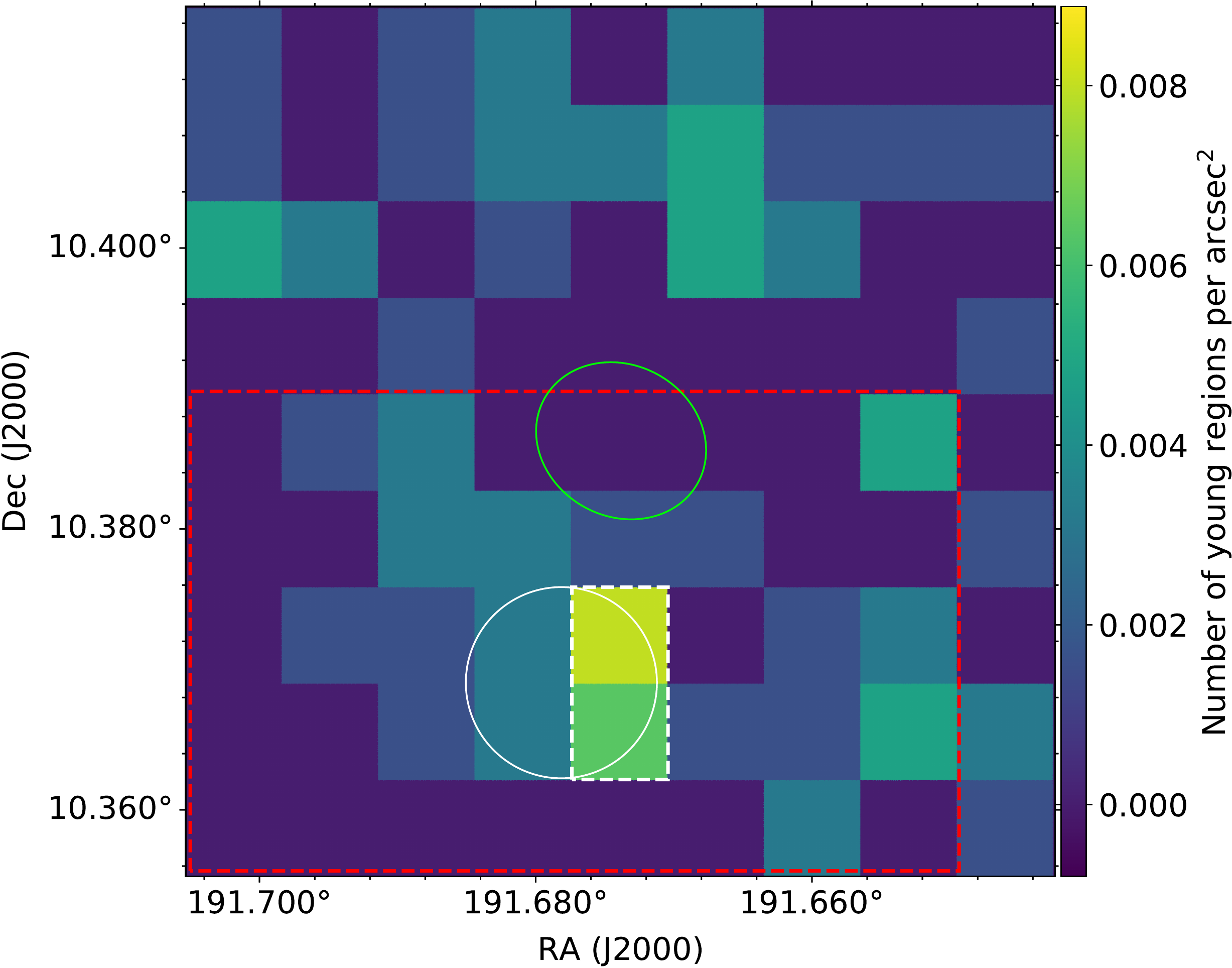}
  \caption{Number density of the \textit{u}-selected blue ($u-g<0.4$ mag) regions around the galaxy \udg{} (marked as the green ellipse). The white circle marks the position of the VLA HI detection of \agc{} from \citet{cannon2015}, with a beam size of 49$\arcsec$. The white dashed box indicates the region above $3\sigma$ level of the mean background number density. The red dashed box ($190\arcsec\times120\arcsec$) shows the area where the  
properties of the blue regions are analyzed in this work.}
     \label{object_density_map}
\end{figure}

To identify newly formed regions among the peak-selected ones, we compared their $u-g$ color to different models for a single burst population of varying ages and metallicities created with Starburst99 \citep{starburst99}.
For this purpose we used the same models adopted in \citet{boselli09} created using Starburst99 models, with a \citet{kroupa01} initial mass function (IMF) between 0.1 and 100 M$_{\odot}$, and four different metallicities (0.05, 0.4, 1, 2 Z$_{\odot}$) based on Geneva stellar evolution tracks. Figure \ref{u-g_color_evolution_starburst99} shows that the $u-g$ colors of single bursts are similar ($u-g \sim 0.4$ mag) when close to 100 Myr, regardless of the metallicity, and bluer colors always correspond to younger regions.
We therefore adopt the limit of $u-g<0.4$ mag in order to be sure to include 
regions dominated by a young stellar population (age < 100 Myr)\footnote{By doing such a color selection, we are aware that we introduce a bias to young regions without taking into to account older regions that could have formed as a result of tidal interactions.}.
Following this color cut, we also removed regions corresponding to known background NED sources.
Unfortunately, this exercise cannot be done using the catalogue of NGVS photometric redshifts from \citet{raichoor2014}, simply because these photo-z were derived using templates not optimized to detect individual HII regions such as those discovered in this work\footnote{For instance, 
three sources in this field, cataloged by \citet{raichoor2014} at $z > 1.75$, 1.67, and 0.44, respectively, have been detected in VESTIGE \Ha{}, and thus are bona fide Virgo cluster objects.}.
The number density of our tentative young $u$-band-selected regions around the galaxy, measured within a grid of $25\arcsec\times25\arcsec$ boxes, is shown in Fig. \ref{object_density_map}. The box size of $25\arcsec$ was chosen to sample the \agc{} HI beam size of $49\arcsec$ from \citet{cannon2015}. We find a clear over-density of young regions south of the galaxy, which coincides with the HI detection of \agc{}. This confirms the visual impression that the blue knots are associated to \agc{}.
In the following, we focus on this side of the galaxy, keeping only regions in the red dashed box in Fig. \ref{object_density_map}, where a total of 38 regions are selected (shown as yellow circular regions in Fig. \ref{region_boxes}).

We estimated the possible contamination of background sources in these \textit{u}-band-selected regions using the object density map shown in Fig. \ref{object_density_map}. We obtain that 29 out of our 38 \textit{u}-band-selected regions fall outside the $3\sigma$ level of the mean background density (white-dashed zone in Fig. \ref{object_density_map}), indicating that we cannot reject the assumption that they are 
background contaminants. However, for the 9 remaining regions forming an over density coinciding with the \agc{} HI detection, we ran a Monte Carlo simulation of a million chains,  estimating that the probability of such an over density being due to contaminants is less than $0.0131\%$.

\subsubsection{Ultraviolet selection}\label{uv_selection}

\begin{figure*}[!ht]
\centering
\includegraphics[width=0.9\hsize]{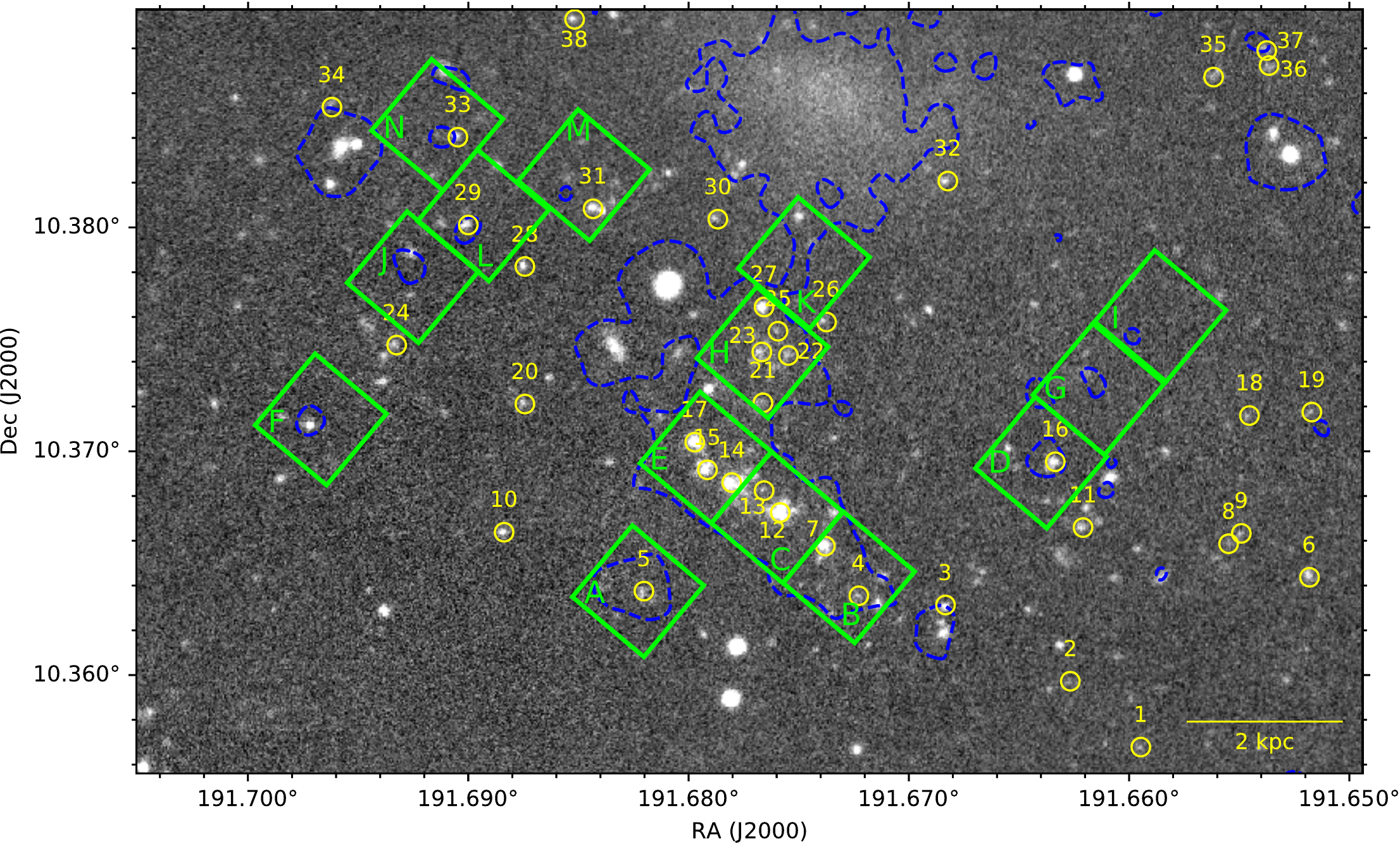}

\includegraphics[width=0.9\hsize]{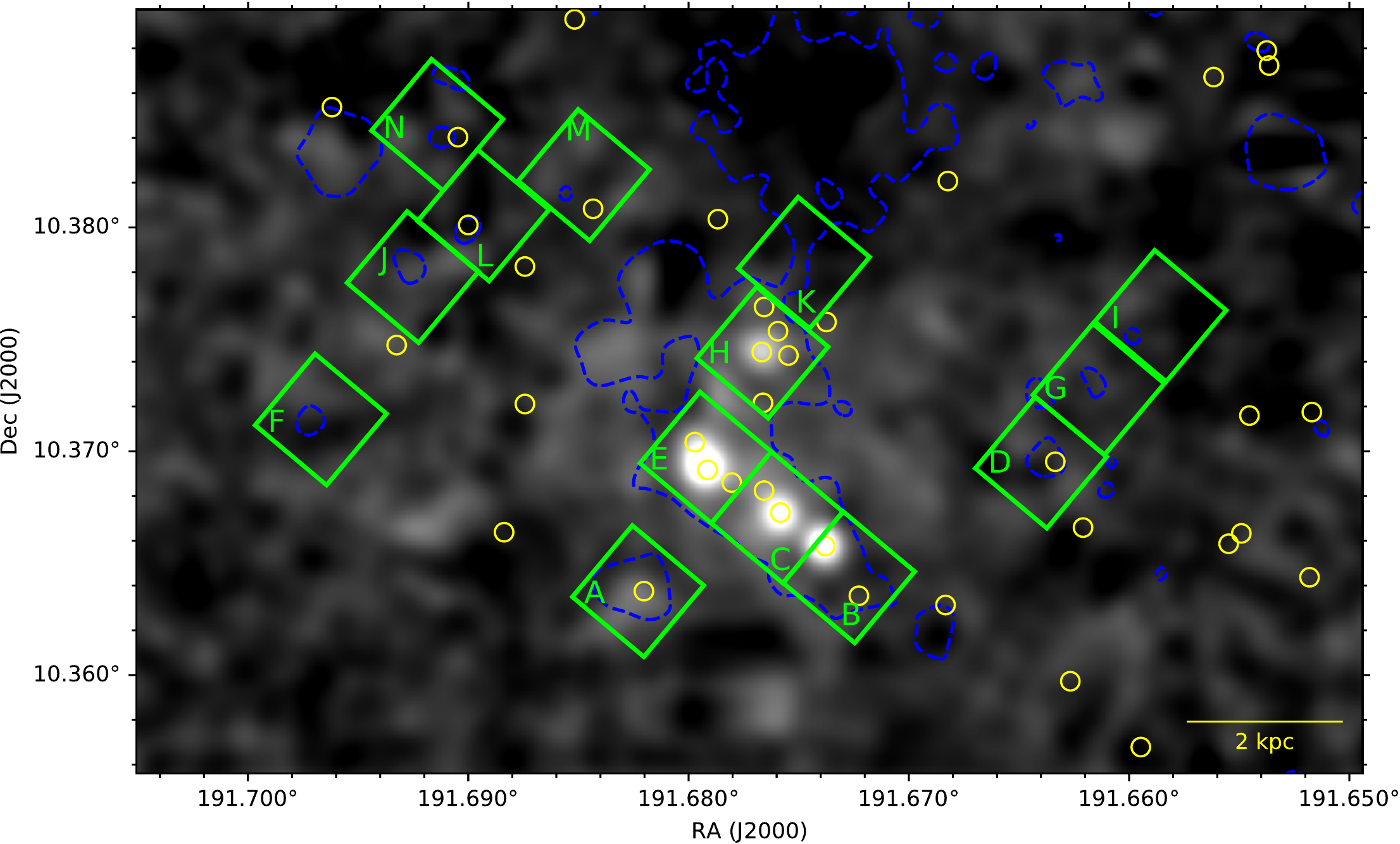}
  \caption{\textit{Top:}\textit{u}-band grayscale image along the area selected for our analysis. \textit{Bottom:} VESTIGE continuum-subtracted \Ha{} image smoothed at the resolution of GALEX. The yellow circles and the green boxes, marked with their names, are respectively our \textit{u}-band-selected and UV-selected regions, as labeled in Table \ref{measurements_table}. The region names are attributed based on increasing declination. The blue dashed lines are the same NUV contours from GALEX as shown in Fig. \ref{color_image}.}
     \label{region_boxes}     
\end{figure*}

Inspection of the NUV image (blue dashed contours in Fig. \ref{region_boxes}) revealed a diffuse emission in the same area or in the vicinity of the blue knot regions. For the sake of completeness, we also made a manual selection based on the UV images (because of the GALEX resolution, we have to work at a lower resolution).
Initially, we created contours on the NUV image of the galaxy at an arbitrarily low surface brightness level of 27 mag arcsec$^{-2}$ after smoothing the data to 3\arcsec resolution (2 GALEX pixels). These contours (shown in Fig. \ref{region_boxes}) were used to visually identify UV-emitting regions associated to the galaxy, shown in Fig. \ref{region_boxes} as green boxes of size $15\arcsec\times15\arcsec$, after excluding any background source identified in NED. A total of 14 regions were finally selected.
Many of them (9 out of 14 regions) also coincide with some of the regions selected in the \textit{u}-band (at higher resolution).
Following \citet{boselli18_vestige3}, we estimated the possible contamination of background UV sources using the number counts given in \citet{Xu2005}. At the limiting magnitude of our detections ($NUV \leq 22.7$ mag, for a GALEX Medium Imaging Survey; \citealt{Morrissey2005}), the expected number of background galaxies is $\sim$1700 sources per deg$^{2}$, or equivalently $\sim$3 for the selected region.
We therefore find a clear excess of UV emission likely related to the galaxy.

The photometric measurements for
these UV-selected young regions were performed similarly to the
\textit{u}-band-selected regions with the {\tt Photutils} \textit{aperture\_photometry} package in all available bands,
after convolving the optical band images with a Gaussian kernel to match the GALEX resolution, and correcting for Milky Way foreground Galactic extinction as discussed in Sect. \ref{profile_measurements_of_udg}.
%

%
\begin{table*}[h!]
\caption{Photometry for the \textit{u}-band selected (top panel) and UV selected regions (bottom panel) as denoted in Fig. \ref{region_boxes}.}
\centering
\small \scalebox{0.9}{\begin{tabular}{cccccccccc}
\hline
\hline
ID  & Distance  &   \textit{u} & \textit{g} & \textit{r} & \textit{i} & \textit{z} & \Ha{} Flux & \textit{NUV} & \textit{FUV} \\  \\
    &   (kpc)   & (mag)    &    (mag) &    (mag)  &   (mag)   &   (mag)   &   (10$^{-16}$ erg s$^{-1}$ cm$^{-2}$)   &   (mag)   &   (mag)   \\
{\tiny(1)}  & {\tiny(2)} & {\tiny(3)} & {\tiny(4)}  & {\tiny(5)}    &   {\tiny(6)}  &   {\tiny(7)}  &   {\tiny(8)}  &   {\tiny(9)}         &   {\tiny(10)}\\
 \hline
 \\
1 & 9.4 & $24.97 \pm 0.21$ & $24.96 \pm 0.22$ & > 24.45 & $24.31 \pm 0.34$ & > 23.95 & < 0.14 & -- & -- \\
2 & 8.3 & $25.26 \pm 0.29$ & $25.3 \pm 0.32$ & > 24.45 & > 24.47 & > 23.95 & < 0.14 & -- & -- \\
3 & 6.8 & $24.1 \pm 0.1$ & $23.84 \pm 0.08$ & $23.85 \pm 0.22$ & $23.47 \pm 0.15$ & $23.42 \pm 0.23$ & < 0.14 & -- & -- \\
4 & 6.6 & $24.52 \pm 0.14$ & $24.34 \pm 0.12$ & $24.06 \pm 0.26$ & $23.78 \pm 0.2$ & > 23.95 & < 0.14 & -- & -- \\
5 & 6.9 & $24.26 \pm 0.11$ & $24.26 \pm 0.12$ & > 24.45 & > 24.47 & > 23.95 & $0.24 \pm 0.02$ & -- & -- \\
6 & 8.9 & $24.04 \pm 0.08$ & $23.82 \pm 0.08$ & $23.92 \pm 0.23$ & $23.22 \pm 0.12$ & $23.22 \pm 0.18$ & < 0.14 & -- & -- \\
7 & 5.9 & $23.01 \pm 0.05$ & $22.94 \pm 0.05$ & $22.44 \pm 0.06$ & $22.46 \pm 0.06$ & $22.38 \pm 0.08$ & $1.54 \pm 0.03$ & -- & -- \\
8 & 7.8 & $25.08 \pm 0.24$ & $25.08 \pm 0.25$ & > 24.45 & $24.3 \pm 0.34$ & > 23.95 & < 0.14 & -- & -- \\
9 & 7.8 & $24.88 \pm 0.2$ & $24.93 \pm 0.22$ & > 24.45 & $24.03 \pm 0.25$ & > 23.95 & < 0.14 & -- & -- \\
10 & 7.0 & $24.1 \pm 0.1$ & $24.02 \pm 0.1$ & $23.48 \pm 0.15$ & $23.41 \pm 0.14$ & > 23.95 & < 0.14 & -- & -- \\
11 & 6.6 & $24.12 \pm 0.1$ & $23.99 \pm 0.09$ & $23.56 \pm 0.16$ & $23.04 \pm 0.1$ & $23.03 \pm 0.16$ & < 0.14 & -- & -- \\
12 & 5.5 & $22.48 \pm 0.05$ & $22.4 \pm 0.05$ & $22.42 \pm 0.06$ & $22.68 \pm 0.07$ & $23.12 \pm 0.18$ & $1.17 \pm 0.02$ & -- & -- \\
13 & 5.2 & $24.58 \pm 0.15$ & $24.54 \pm 0.15$ & > 24.45 & > 24.47 & > 23.95 & $0.27 \pm 0.02$ & -- & -- \\
14 & 5.2 & $22.63 \pm 0.05$ & $22.57 \pm 0.05$ & $22.74 \pm 0.08$ & $22.9 \pm 0.08$ & $23.27 \pm 0.2$ & $0.2 \pm 0.02$ & -- & -- \\
15 & 5.1 & $22.9 \pm 0.05$ & $22.95 \pm 0.05$ & $22.9 \pm 0.08$ & $23.06 \pm 0.1$ & $23.84 \pm 0.36$ & $1.59 \pm 0.03$ & -- & -- \\
16 & 5.7 & $23.51 \pm 0.05$ & $23.27 \pm 0.05$ & $23.02 \pm 0.1$ & $22.49 \pm 0.06$ & $22.32 \pm 0.08$ & < 0.14 & -- & -- \\
17 & 4.9 & $22.85 \pm 0.05$ & $22.55 \pm 0.05$ & $22.32 \pm 0.05$ & $22.4 \pm 0.06$ & $22.64 \pm 0.11$ & $0.55 \pm 0.02$ & -- & -- \\
18 & 6.9 & $24.92 \pm 0.2$ & $24.85 \pm 0.2$ & > 24.45 & $24.18 \pm 0.3$ & > 23.95 & < 0.14 & -- & -- \\
19 & 7.5 & $24.98 \pm 0.21$ & $25.06 \pm 0.25$ & > 24.45 & > 24.47 & > 23.95 & < 0.14 & -- & -- \\
20 & 5.6 & $24.47 \pm 0.13$ & $24.24 \pm 0.12$ & $23.88 \pm 0.22$ & $23.52 \pm 0.16$ & $23.69 \pm 0.3$ & < 0.14 & -- & -- \\
21 & 4.1 & $24.32 \pm 0.12$ & $24.08 \pm 0.1$ & $23.87 \pm 0.22$ & $23.06 \pm 0.1$ & $22.86 \pm 0.14$ & < 0.14 & -- & -- \\
22 & 3.5 & $24.09 \pm 0.09$ & $23.96 \pm 0.08$ & $23.77 \pm 0.2$ & $23.57 \pm 0.16$ & $23.44 \pm 0.24$ & < 0.14 & -- & -- \\
23 & 3.5 & $23.74 \pm 0.06$ & $23.71 \pm 0.07$ & $23.53 \pm 0.16$ & $23.74 \pm 0.19$ & $23.84 \pm 0.36$ & $0.64 \pm 0.02$ & -- & -- \\
24 & 6.4 & $24.4 \pm 0.12$ & $24.28 \pm 0.12$ & $23.7 \pm 0.18$ & $23.26 \pm 0.12$ & > 23.95 & < 0.14 & -- & -- \\
25 & 3.2 & $24.76 \pm 0.17$ & $24.87 \pm 0.21$ & > 24.45 & $24.16 \pm 0.29$ & > 23.95 & $0.2 \pm 0.02$ & -- & -- \\
26 & 3.0 & $24.53 \pm 0.14$ & $24.37 \pm 0.13$ & $24.1 \pm 0.28$ & $24.08 \pm 0.26$ & $23.62 \pm 0.29$ & < 0.14 & -- & -- \\
27 & 2.9 & $23.14 \pm 0.05$ & $23.14 \pm 0.05$ & $22.96 \pm 0.1$ & $22.55 \pm 0.06$ & $22.6 \pm 0.11$ & < 0.14 & -- & -- \\
28 & 4.5 & $24.24 \pm 0.11$ & $24.17 \pm 0.11$ & $24.27 \pm 0.33$ & $23.98 \pm 0.24$ & > 23.95 & < 0.14 & -- & -- \\
29 & 4.9 & $23.9 \pm 0.08$ & $23.7 \pm 0.07$ & $23.34 \pm 0.13$ & $22.74 \pm 0.08$ & $22.73 \pm 0.12$ & < 0.14 & -- & -- \\
30 & 2.2 & $24.52 \pm 0.14$ & $24.32 \pm 0.12$ & $24.06 \pm 0.26$ & $23.52 \pm 0.16$ & $23.46 \pm 0.24$ & < 0.14 & -- & -- \\
31 & 3.4 & $23.52 \pm 0.06$ & $23.41 \pm 0.05$ & $23.08 \pm 0.11$ & $23.32 \pm 0.12$ & $23.14 \pm 0.18$ & < 0.14 & -- & -- \\
32 & 2.0 & $24.08 \pm 0.09$ & $23.9 \pm 0.08$ & $23.46 \pm 0.15$ & $23.38 \pm 0.14$ & $23.0 \pm 0.15$ & < 0.14 & -- & -- \\
33 & 4.8 & $24.1 \pm 0.1$ & $24.21 \pm 0.11$ & $23.92 \pm 0.23$ & $23.58 \pm 0.16$ & $23.64 \pm 0.29$ & < 0.14 & -- & -- \\
34 & 6.3 & $24.72 \pm 0.16$ & $25.03 \pm 0.24$ & $24.15 \pm 0.29$ & $24.12 \pm 0.28$ & > 23.95 & < 0.14 & -- & -- \\
35 & 5.0 & $24.46 \pm 0.13$ & $24.31 \pm 0.12$ & > 24.45 & $24.16 \pm 0.29$ & $23.87 \pm 0.37$ & < 0.14 & -- & -- \\
36 & 5.7 & $24.42 \pm 0.12$ & $24.34 \pm 0.12$ & $23.37 \pm 0.14$ & $22.95 \pm 0.09$ & $22.55 \pm 0.1$ & < 0.14 & -- & -- \\
37 & 5.7 & $24.38 \pm 0.12$ & $24.2 \pm 0.11$ & $24.22 \pm 0.32$ & $23.64 \pm 0.17$ & $23.22 \pm 0.19$ & < 0.14 & -- & -- \\
38 & 3.3 & $24.16 \pm 0.1$ & $23.96 \pm 0.08$ & $23.28 \pm 0.12$ & $22.78 \pm 0.08$ & $22.56 \pm 0.11$ & < 0.14 & -- & -- \\
\\
\hline\\

A & 6.9 & > 22.29 & > 22.21 & > 21.64 & > 21.26 & > 20.8 & < 1.79 & $22.56 \pm 0.12$ & $22.48 \pm 0.11$ \\
B & 6.3 & $21.82 \pm 0.24$ & $21.2 \pm 0.15$ & $20.7 \pm 0.16$ & $20.0 \pm 0.12$ & $20.26 \pm 0.24$ & $2.82 \pm 0.27$ & $21.81 \pm 0.06$ & $21.81 \pm 0.06$ \\
C & 5.6 & $20.94 \pm 0.11$ & $20.74 \pm 0.1$ & $20.86 \pm 0.18$ & $20.48 \pm 0.18$ & > 20.8 & $5.45 \pm 0.26$ & $20.89 \pm 0.05$ & $20.87 \pm 0.05$ \\
D & 5.6 & > 22.29 & $22.0 \pm 0.32$ & > 21.64 & $21.16 \pm 0.36$ & > 20.8 & < 1.79 & $23.26 \pm 0.22$ & > 23.82 \\
E & 5.0 & $20.94 \pm 0.11$ & $20.68 \pm 0.08$ & $20.69 \pm 0.15$ & $20.22 \pm 0.14$ & $20.61 \pm 0.33$ & $5.63 \pm 0.26$ & $20.79 \pm 0.05$ & $20.83 \pm 0.05$ \\
F & 7.8 & > 22.29 & > 22.21 & > 21.64 & > 21.26 & > 20.8 & < 1.79 & $23.7 \pm 0.36$ & > 23.82 \\
G & 5.2 & > 22.29 & > 22.21 & > 21.64 & $20.59 \pm 0.2$ & $20.42 \pm 0.26$ & < 1.79 & $23.52 \pm 0.29$ & > 23.82 \\
H & 3.5 & $21.42 \pm 0.16$ & $21.2 \pm 0.14$ & $20.63 \pm 0.15$ & $19.78 \pm 0.1$ & $19.62 \pm 0.12$ & $3.16 \pm 0.27$ & $21.48 \pm 0.05$ & $21.64 \pm 0.05$ \\
I & 5.2 & > 22.29 & > 22.21 & > 21.64 & $20.97 \pm 0.3$ & $20.66 \pm 0.34$ & < 1.79 & > 23.82 & > 23.82 \\
J & 5.8 & > 22.29 & > 22.21 & > 21.64 & $20.86 \pm 0.26$ & > 20.8 & < 1.79 & > 23.82 & > 23.82 \\
K & 2.3 & $22.06 \pm 0.32$ & $21.4 \pm 0.18$ & $20.78 \pm 0.16$ & $20.12 \pm 0.13$ & $20.2 \pm 0.22$ & < 1.79 & $22.92 \pm 0.16$ & $23.26 \pm 0.22$ \\
L & 4.7 & > 22.29 & > 22.21 & > 21.64 & $21.17 \pm 0.37$ & > 20.8 & < 1.79 & > 23.82 & > 23.82 \\
M & 3.3 & > 22.29 & > 22.21 & > 21.64 & > 21.26 & > 20.8 & < 1.79 & > 23.82 & > 23.82 \\
N & 5.0 & > 22.29 & > 22.21 & > 21.64 & > 21.26 & > 20.8 & < 1.79 & $23.66 \pm 0.34$ & > 23.82 \\

\\
\hline
\end{tabular}}
\tablefoot{(1) Name of the region. (2) Projected distance of the region from the center of \udg{}. (3-7) \textit{u}, \textit{g}, \textit{r}, \textit{i} and \textit{z}-band magnitudes. (8) VESTIGE \Ha{} flux. (9-10) GALEX \textit{NUV} and \textit{FUV} magnitudes. The upper limits ($3\sigma$) in the broad-band magnitudes and in the \Ha{} fluxes are denoted with > and < symbols, respectively.}
\label{measurements_table}
\end{table*}
%

\section{Analysis}\label{analysis}

\subsection{Galaxy evolution models with ram-pressure stripping applied to \udg{}}
\label{analysis_rps}

In previous studies,  \citet{boselli06, boselli08, boselli08_2, boselli14_guvics} reproduced the properties of anaemic and dwarf galaxies located in the Virgo Cluster by adding RPS to chemical and spectrophotometric evolution models initially made for unperturbed galaxies. These models were first  developed for the Milky Way and for nearby spirals \citep{boissier99,boissier2000,munos2011}. Their output are radial profiles of stellar density, mass density, metallicity, and spectra. 
The models are constructed making some assumptions on the final total mass distribution within the disk (in the absence of interactions that remove gas) and the surrounding halo, the gas accretion history, the \citet{kroupa01} IMF, and the star formation law.
They were calibrated in such a way that the only two free parameters (for the unperturbed case) are the spin ($\lambda$, specific angular momentum) and the rotational velocity ($V$), which is tightly connected to the total mass of the galaxies ($M \propto V^3$).
The same models have been adapted to reproduce the evolution of LSB galaxies \citep{boissierlsb} by assuming large spin parameters, as commonly done in the literature \citep{jimenez1998,amorisco2016}.

However, a full grid including both LSB (i.e., large and very large spins) and RPS has not been computed so far. For the study of the sample that will be presented in Junais et al. (in preparation), we prepared a very large grid with the same models, but covering a very large range of spin parameters (from 0.01 to 0.6 in steps of 0.01), in order to include the spin corresponding to the very extended disk of Malin 1 \citep{boissier16}, and of velocity (from low-mass dwarf galaxies with $V=20$ km s$^{-1}$ to very massive galaxies with $V=600$ km s$^{-1}$, with steps of 2 and 10 km s$^{-1}$ , respectively, below and above 150 km s$^{-1}$ to better sample the low-mass range in which galaxies are more numerous). 
The ram-pressure stripping event was modeled as described in \citet{boselli06}. 
In practice, we remove gas at a rate of $\epsilon \Sigma_{gas}/ \Sigma_{potential}$, which is proportional to the galaxy gas column density at any given time but is modulated by the potential of the galaxy, measured
by the total (baryonic) local density. $\epsilon$ is linked to the RPS efficiency and follows a Gaussian with a maximum value $\epsilon_0$
at the peak time ($t_{rps}$), assuming that the current age of the galaxy is 13.5 Gyr. 
This time variation was chosen to mimic that obtained by \citet{vollmer2001} for a galaxy crossing the Virgo cluster potential on an elliptical orbit.
To reduce the number of free parameters ($\lambda$, $V$, $t_{rps}$) we  keep the same peak efficiency ($\epsilon_0$) of 1.2 $M_{\odot}$ kpc$^{-2}$ yr$^{-1}$, and the average FWHM of the Gaussian variation from  \citet{vollmer2001} of $\simeq$150 Myr, as in \citet{boselli06}.
We included various $t_{rps}$ values from 8 (distant past) to 13.6 Gyr (for which the peak of RPS will occur 0.1 Gyr in the future), with steps of 0.1 Gyr (considering the timescale of the various processes involved, including ram pressure, the models are not sensitive to much shorter times).

The fact that we keep a constant peak efficiency and a unique FWHM is clearly an over-simplification of the problem. Indeed these parameters should depend on the precise orbit within the cluster. However, we choose  to do it as it allows us to explore a large grid of models for the other parameters, within reasonable computational time. This grid will also be used for a study of about 150 low-surface-brightness galaxies in the Virgo Cluster (Junais et al., in preparation) for which we cannot fine-tune the orbit parameters. However, below, we discuss  the uncertainties that this assumption brings to the properties derived in the present paper. 

\begin{figure}
\centering
\includegraphics[width=\hsize]{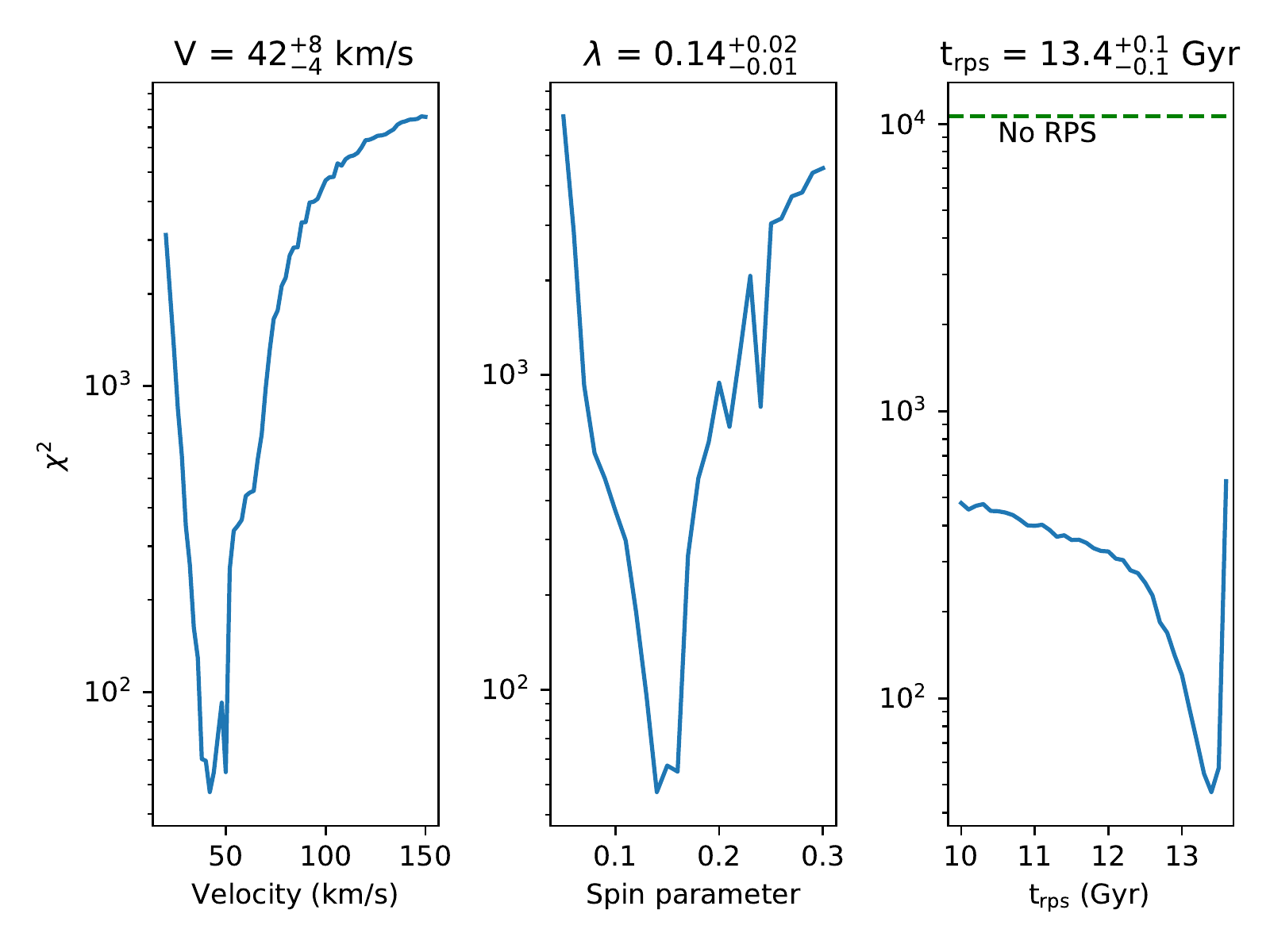}
  \caption{\chisq{} distribution for the determination of the best model parameters ($V$, $\lambda$ and $t_{\mathrm{rps}}$) for \udg{}, as discussed in Sect. \ref{analysis_rps}. The best-fit parameters 
  are given in the upper part of each panel. The given confidence limits (3$\sigma$) in each parameter are obtained following \citet{avni}. The green horizontal dashed line in the right panel marks the \chisq{} value corresponding to a model without RPS, keeping the same values of $V$ and $\lambda$ as in the best-fit model.}
    \label{models_chisq}
\end{figure}

Figure \ref{models_chisq} shows the \chisq{} distribution around our best solution. We note that we computed values of \chisq{}, adopting a minimum error of 0.05 mag to take into account systematic uncertainties (e.g., IMF, stellar tracks, stellar libraries). We rejected any solution violating the $3\sigma$ upper limits of our photometry. However, we kept a tolerance of 0.1 mag again to take into account systematic uncertainties in the stellar population models. We found that this helped us to avoid rejecting a good  model that only marginally violates one upper limit. Modifying this tolerance within a range of a few tenths of dex changes the best-fit parameters within their error bars.

The best model was obtained for a ram-pressure stripping event peaking 100 Myr ago ($t_{rps}=13.4$ Gyr) in a low-mass galaxy ($V=42$ km s$^{-1}$) with a large spin ($\lambda=0.14$). This solution is much better than any model without RPS, as shown in Fig. \ref{models_chisq}. Values of spin around $\lambda=0.14$ were already found to reproduce LSB galaxies in \citet{boissierlsb}, but on average for more massive and brighter galaxies than our UDG (with $V$ in the range 40-360 km s$^{-1}$ instead of 42 km s$^{-1}$; and absolute magnitudes $M_B$ typically in the range -14 to -22 mag, while the UDG in this paper has an absolute $g$-band magnitude of -13.6 mag). The low velocity of 42 km s$^{-1}$ obtained from our best model reasonably matches the properties of dwarf galaxies in Virgo \citep{boselli08,boselli08_2}, but with a more typical spin of $\lambda=0.05$. This means that only the combination of a dwarf galaxy velocity and a large spin allows the model to reproduce the properties of \udg{}. Cosmological simulations indicate that the spin parameter distribution is expected to be lognormal, with a peak at 0.05, and $\sigma=0.5$ \citep{mo1998}, but cold gas accretion could lead to larger angular momentum \citep{stewart2017}. The value of $\lambda=0.14$ is beyond the peak of the distribution, but is not totally unexpected based on these considerations.

While our best model was obtained by keeping only three free parameters, we now investigate how our results are affected by the peak efficiency ($\epsilon_0$) and the FWHM of the Gaussian used for modelling the RPS event. For this, we decided to keep a constant spin and velocity, because these two parameters, affecting mostly the long-wavelength range, are weakly affected by a recent RPS event (in \citealt{boselli06}, they were chosen on the basis of the H-band profile alone, and the rotation curve). We then computed models with $\epsilon_0$ in the range of 0.2 to 1.6 $M_{\odot}$ kpc$^{-2}$ yr$^{-1}$ that was considered in \citet{boselli06}, and $t_{rps}$ in the range of 13.0 to 13.6 Gyr (because our best fit clearly indicates a recent RPS event). We kept any of these models with $\chi^2$ lower than the limit considered above, and not violating upper limits. In another test, we kept the efficiency fixed to 1.2 $M_{\odot}$ kpc$^{-2}$ yr$^{-1}$ as in the original model, but allowed the FWHM of the Gaussian shape of the RPS event to vary within the range of 100 to 200 Myr (in steps of 10 Myr) as presented by \citet{vollmer2001}, and carried out the same procedure. We thus obtained several models consistent with the data, for various $\epsilon_0$ and FWHM. For these models the best $t_{rps}$ is found to be 13.3 or 13.4 Gyr. The obtained profiles are within the red-shaded region shown in Fig. \ref{profiles_and_model}. They are very similar to the best model derived above, except for \Ha{} in which we obtain a larger dispersion among models due to
small values of $\epsilon_0$ allowing the galaxy to keep more gas, and $t_{rps}=13.3$ leaving more time for some gas to return from old stars after the RPS event. The gas removal still allows the models to be within the observed \Ha{} upper limit. These models allow us to estimate the uncertainty we introduce in the quantities we derive by fixing the RPS model parameters.

\begin{figure}

\centering
\includegraphics[width=\hsize]{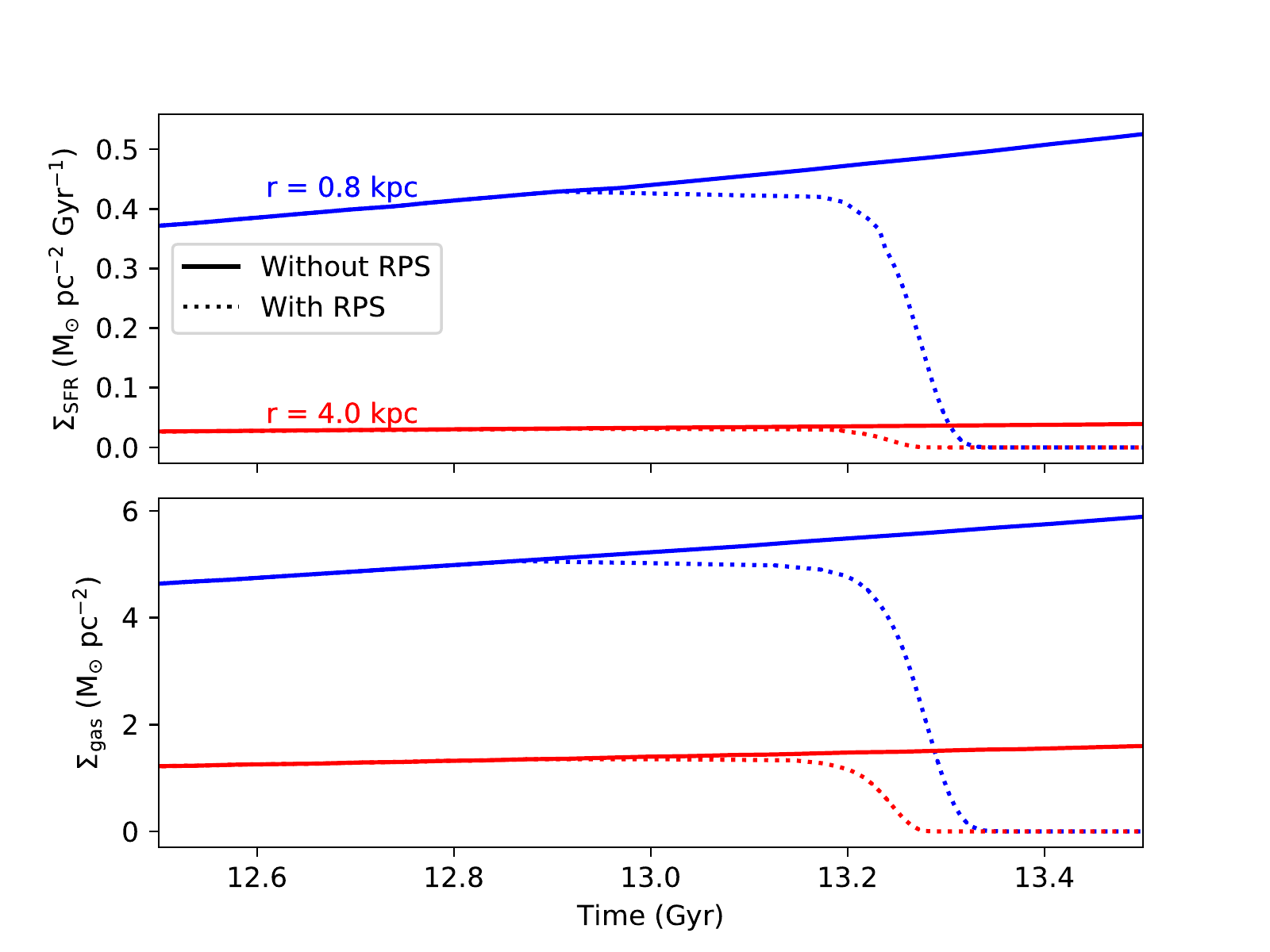}
  \caption{Evolution of the surface densities of SFR and gas for the disk of \udg{} obtained in our models. The dotted and solid lines show the models with and without RPS, respectively. The blue and red curves show the evolution at 0.8 kpc and 4 kpc from the galaxy center.}
    \label{plot_sfrd_sigma_gas_time}
\end{figure}

The profile of the best model is shown as the black dotted line in Fig. \ref{profiles_and_model}. The multi-wavelength profiles (including upper limits) of the UDG galaxy are very well fitted by these RPS models, except for FUV, in which the best-fit model underpredicts the observed FUV central detection by $\sim$0.5 mag. Such a difference can be attributed, for example, to our assumption on the IMF \citep{munos2011}, or the adopted stellar spectra library (more uncertain in the UV than the optical). Moreover, our previous tests show that the UV surface brightness is also sensitive to the precise value of FWHM and $\epsilon_0$ in the RPS models, as can be seen in Fig. \ref{profiles_and_model}. We also show the profile for a model with the same velocity and spin, but without any ram pressure (i.e., what would have happened to the galaxy in the absence of RPS). We can clearly see that the \Ha{} upper limits and UV data are of paramount importance to show that ram pressure was recently present. Indeed, on short timescales, only these bands are very sensitive to the gas removal and quenching of star formation.
Figure \ref{plot_sfrd_sigma_gas_time} illustrates these phenomena at two different radii ($R=$ 0.8 and 4.0 kpc), showing the evolution of the gas and star formation rate surface density with time for the best model and the same model without RPS. While the peak of RPS occurred 100 Myr ago, the gas removal and the quenching of the star formation began a few hundred million years before, when the galaxy was first entering into the cluster. Due to the shallow gravitational potential well of this UDG, the gas stripping process was very efficient well before the galaxy reached the core of the cluster (200 Myr ago).
While the efficiency of ram pressure evolves as a Gaussian, the gas-loss rate is not symmetric around the peak because most of the gas has 
already been removed at that time. This is similar to what was found with much more sophisticated models of RPS by \citet{roediger2005}.

\begin{table}
\caption{Properties of the best RPS models, and the model with the same spin velocity but without the RPS.}             
\centering                          
\scalebox{1}{\begin{tabular}{l c c}        
\hline\hline                 
Property & RPS models  & Model without RPS\\    
\hline                        
$R_{e,g}$ (kpc) & $1.98 \pm 0.03 $ & $2.23 \pm 0.52 $ \\
$\mu_{0,g}$ (\magperarcsec{}) & $25.25 \pm 0.08 $ & $24.61 \pm 0.59 $ \\
$\log$ \mstar{} (\msun) & $7.10 \pm 0.02 $ & $7.31 \pm 0.27 $ \\
$\log M_{\mathrm{gas}}$ (\msun) & $5.35 \pm 0.42 $ & $8.60 \pm 0.17 $ \\
$\log$ SFR (\msun yr$^{-1}$) & $-5.80 \pm 0.43 $ & $-1.86 \pm 0.25 $ \\
$Z_{\mathrm{gas}}$ ($Z_{\odot}$)  & $0.38 \pm 0.05 $ & $0.16 \pm 0.05 $ \\
\hline                                   
\end{tabular}}
\label{galaxymodel_table}
\end{table}


Finally, several global properties of these models are given in Table \ref{galaxymodel_table}. For the RPS models, the table provides the average value and the range obtained for the models satisfying our criteria among all those tested with various $\epsilon_0$ and FWHM for the RPS event (removing a couple of outlying metallicity values, see below). For the unperturbed model, we indicate the dispersion obtained when considering 
the uncertainties on the spin and velocity obtained during the fitting process. 
The stellar mass, central surface brightness, and effective radius are more affected by the uncertainties on the spin and velocity than by the RPS choices. The uncertainties on the gas left and the SFR in the RPS models are on the contrary dominated by the RPS choices. When plotting these values for the RPS models, we combine the errors due to RPS assumptions and the one related to the fitting of the spin and velocity.
Finally, the table does not include the systematic effects that have to be kept in mind, such as the fact that stellar masses are dependent on the IMF, and metallicities are dependent on the yields adopted in the models, implying an uncertainty of about a factor of two in both the cases.

The comparison of the RPS models  with the unperturbed one also tells us how much the galaxy is affected by the ongoing RPS event (indeed, the galaxy before the RPS event was almost in the same state as the nonRPS model considering the timescales involved).
While the unperturbed galaxy was dominated by the gas \citep[a standard result for models with low mass and large spin in the context of these models;][]{boissier01}, because of the weak potential of the galaxy, most of the gas has been removed in the RPS model. Star formation has been almost totally quenched with respect to the nonRPS model, consistent with the faintness of the galaxy at UV and blue optical wavelengths.
The gas-phase metallicity is larger in the RPS model. This is to be expected because the metals expelled now by a previous generation of stars reaching the end of their life are diluted in a much smaller amount of remaining gas, as indeed generally observed in gas-poor cluster galaxies \citep{boselli08,hughes13}.
The difference is around a factor two, and the gas-phase metallicity of the unperturbed galaxy was around one-tenth solar. However, we note that the gas-phase metallicity for RPS models, with very small gas fractions, becomes unreliable as it becomes dominated by the yield of the stars dying at that time, which in some cases leads to artificially high values. The corresponding small amount of gas makes this metallicity impossible to observationally measure in any case.

Finally, we note that in the framework of our models, the effective radius has not changed much  and the central surface brightness is dimmer than the nonRPS model by $\sim$0.7 magnitude.  
With these values, the unperturbed galaxy would still be a UDG, but a star-forming one with bluer colors. This could correspond to the blue UDGs for which evidence of existence, especially in the field, has been found by \citet{prole2019}.

A caveat of the models used for this study is that they do not take into account other effects that may modify the effective radius and the central (or effective) surface brightness, such as tidal interaction or adiabatic expansion. However, the properties of the UDGs found in the RomulusC Galaxy Cluster simulation by \citet{tremmel2020} are  mostly determined by RPS (with passive evolution after a quenching event), while tidal interactions play a modest role.
Another caveat is that the models do not take into account some effects that have been proposed as the origin of UDG galaxies, and that can be included in hydro-dynamical modelling, such as for instance very efficient early feedback \citep{martin2019,dicintio2019}. These effects could explain why
the galaxy needs a large angular momentum before the RPS interaction (the galaxy is extended early on, in a way that is not taken into account explicitly in our case, but that we mimic by adopting a large spin). The RPS event is then crucial in quenching star formation and turning the previously blue UDG into a red UDG like the the ones typically found in clusters.

\subsection{Stellar mass and age of the blue knots}\label{age_stellar_mass_determination}

In order to better understand and characterize the nature of the regions selected around \udg{}, we estimated the stellar mass and age of each region based on the photometric measurements given in Table \ref{measurements_table}. To this aim we used the single burst Starburst99 models discussed in Sect. \ref{u_band_selection}. For each of the four metallicities, we performed  
a \chisq{} minimization to find the age and stellar mass providing the best fit to our measurements. 
We considered an arbitrary stellar mass range of $10$ to $10^{7}$ \msun{}, with a spacing of 0.04 in log. For each value of the stellar mass, metallicity, and age, we first checked if the model was violating any of our upper limits shown in Table \ref{measurements_table}. If this was the case, it was rejected. The \Ha{} measurements play a major role in constraining the age of the regions, with an upper limit indicating an age of greater than 10 Myr in massive regions. 
\begin{figure}
\centering
\includegraphics[width=\hsize]{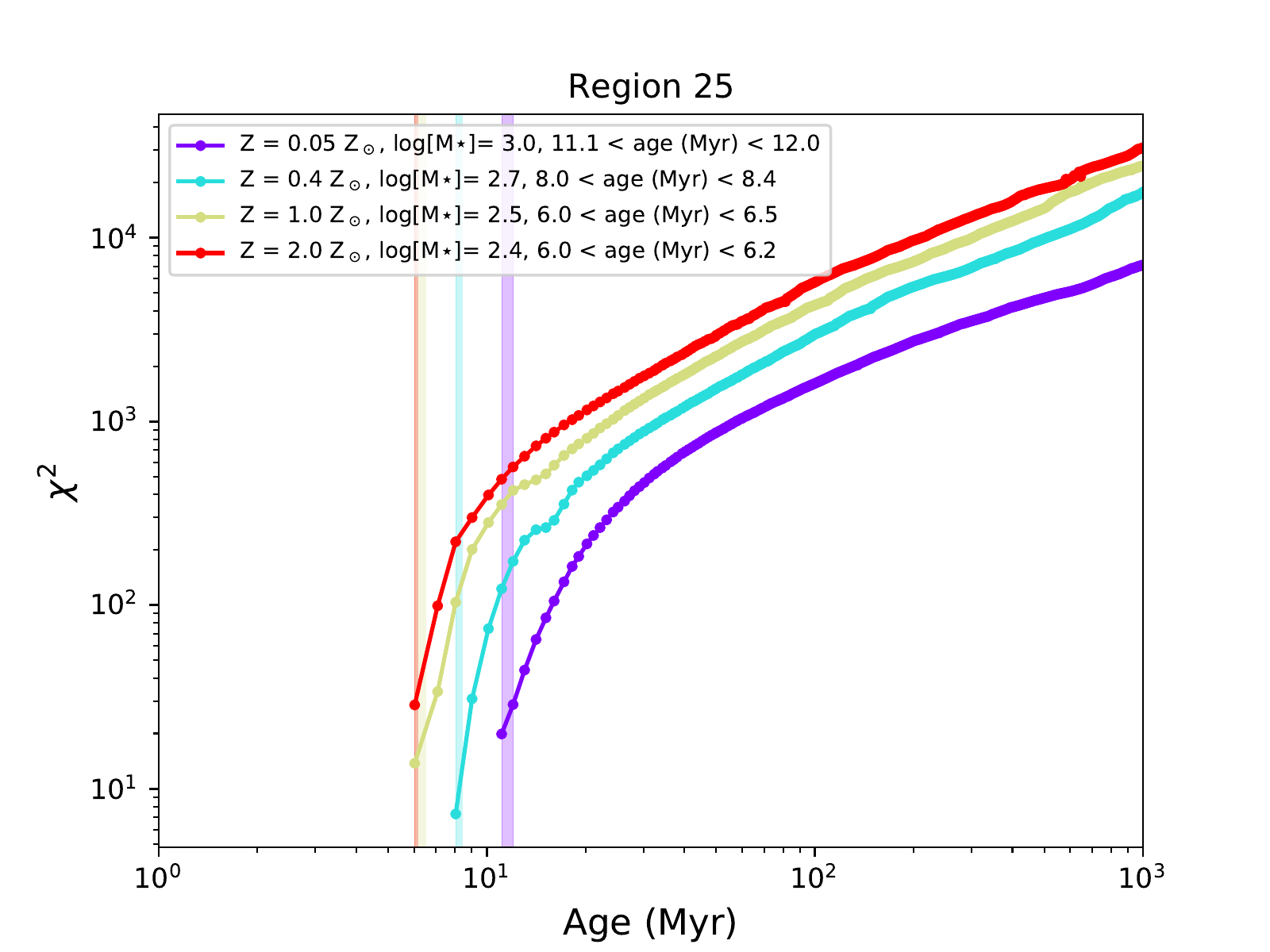}
  \caption{Example of the \chisq{} distribution for determination of the stellar mass, the age and the metallicity of region 25. The confidence limits around the best-fit ages ($3\sigma$), indicated by the vertical bands, are obtained from the given \chisq{} distribution following \citet{avni}. We note that all the models with an age of less than 6 Myr for this region are rejected due to upper limit violations, resulting in the unusual single-sided \chisq{} distribution above (the other side could be represented by a vertical line if we assign artificially infinite \chisq{} to rejected models). Similar \chisq{} distributions for all the other regions are provided in Appendix \ref{appendix:stellar_mass_age_chi2}.}
    \label{chi2region_example}
\end{figure} 
The best \chisq{} are shown as a function of age and metallicity for an example region in Fig. \ref{chi2region_example}. 
The uncertainties on the age are computed from the \chisq{} distribution following \citet{avni}, with a $3\sigma$ confidence level.
We adopt the metallicity providing the least \chisq{}. Ideally, we would expect the metallicity of the regions to be similar to the metallicity in the gas of the galaxy before the stripping event, that is, at 0.16 $Z_{\odot}$, which could be tested with our results. However, in most cases, the lack of data in the observed SEDs prevents us from discriminating between different metallicities (see Fig. \ref{chi2region_example} and Appendix \ref{appendix:stellar_mass_age_chi2} for all the regions). 
As a result, we cannot constrain the metallicity.
We note that the age and stellar mass usually change by less than a few million years and a few tenths of dex, respectively,  over various metallicities. However, with the lowest metallicity ($Z=0.05\,Z_\odot$), it is sometimes possible to obtain older ages (a few 100 Myr) and (up
to ten times) larger  stellar masses  than for the other metallicities, but we only obtain a metallicity of $0.05\,Z_\odot$  for a few of our regions. 

The results of the above procedure for all the regions are given in Table \ref{age_stellarmass_table}. For a few regions, we could only obtain an upper limit or lower limit on age, because our models only cover an evolution within a time range of 1 Myr to 1 Gyr ($\Delta t = 1$ Myr) or the \chisq{} curve does not favor a better constraint. We note that some of the regions have extremely small \chisq{} values. This is  due to the fact that these regions have upper limits in most of the bands, which were not used in the \chisq{} computation. For instance, regions F, J, L, and N have only one measurement that is not an upper limit. In such cases, it is possible to fine-tune each of the three parameters (age, stellar mass, and metallicity) to go exactly through this point, but the resulting model is not really constrained. 
However, even then, we can often put a limit on the age of the burst as upper limits would be violated outside the considered range. For region M, we only have upper limits on photometry because the NUV detection observed in this region is too small for the aperture size we used for the photometry (see Fig. \ref{region_boxes}). Therefore we were not able to perform our \chisq{} minimization procedure on this region.  

Many sources of uncertainty may affect these results, such as for example the minimization procedure (\chisq{} vs. maximum likelihood) and the way upper limits are dealt with, the details of the stellar population models, and our assumption of the absence of dust, or a star formation event (single population vs. extended). While in the remainder of the paper, we use the results presented above, we performed two further tests, which are described below.

i) We compared the color evolution found by \citet{boselli18_vestige3} for bursts with extended star formation histories while we adopted single bursts. The color evolution of \citet{boselli18_vestige3} is always  within the range of colors found in our single-burst models for different metallicities. These could not be distinguished on the basis of our data. Thus, extended bursts or a single population lead to similar results that cannot be distinguished with the data in hand.

ii) We  determined the age and stellar mass of each region
with the  SED-fitting code CIGALE\footnote{\url{https://cigale.lam.fr}} \citep{burgarella05,Noll09,boquien19}, which uses a Bayesian approach and implements different assumptions concerning the sources of uncertainties discussed above. While results for individual regions may vary (with, on average,  older ages in CIGALE, with large error bars),  with both approaches we find several regions with young ages ($<20$ Myr), especially for the regions that are detected in \Ha. This again shows that \Ha{} measurements allowed by the VESTIGE survey are crucial for studying the very young star forming regions studied in this work. A longer discussion of the results obtained with CIGALE and the properties obtained with Starburst99 is given in Appendix \ref{appendix:cigale_fitting}. 

Our tests suggest that although individual values may differ, the existence of young regions with recent star formation is a robust result from our analysis.

%
\begin{table}
\caption{Age, stellar mass, metallicity, and reduced \chisq{} values determined for our selected regions as discussed in Sect. \ref{age_stellar_mass_determination}.} 
\centering                          
\scalebox{0.94}{
\begin{tabular}{c c c c c}        
\hline\hline                 
ID & $M_\star$ & Age & Metallicity & $\chi^2_{red}$ \\
    &  ($10^{3}$ \msun)  &     (Myr)    &   ($Z_\odot$)     &   \\
\hline
\\
1 & 0.8 & $13 \pm 4 $ & 1.00 & 0.0531 \\
2 & 0.7 & $14 \pm 6 $ & 2.00 & 0.0054 \\
3 & 2.8 & $14 \pm 1 $ & 1.00 & 0.4512 \\
4 & 1.4 & $13 \pm 2 $ & 0.40 & 0.0321 \\
5 & 0.5 & $6 \pm 1 $ & 1.00 & 0.3271 \\
6 & 2.8 & $13 \pm 1 $ & 1.00 & 2.1775 \\
7 & 1.6 & $7 \pm 1 $ & 0.40 & 64.3600 \\
8 & 1.1 & $19 \pm 6 $ & 0.40 & 0.0818 \\
9 & 1.3 & $14 \pm 1 $ & 0.40 & 0.2949 \\
10 & 2.8 & $13 \pm 1 $ & 0.40 & 0.8093 \\
11 & 2.5 & $12 \pm 1 $ & 0.40 & 1.7665 \\
12 & 2.3 & $6 \pm 1 $ & 1.00 & 16.5120 \\
13 & 0.9 & $10 \pm 1 $ & 0.05 & 0.5585 \\
14 & 10.0 & $19 \pm 1 $ & 0.05 & 2.3020 \\
15 & 1.1 & $6 \pm 1 $ & 0.40 & 20.2600 \\
16 & 4.4 & $12 \pm 1 $ & 0.40 & 2.6175 \\
17 & 2.1 & $7 \pm 1 $ & 1.00 & 29.1400 \\
18 & 1.1 & $14 \pm 2 $ & 0.40 & 0.0632 \\
19 & 0.8 & $18 \pm 6 $ & 0.05 & 0.0026 \\
20 & 1.0 & $8 \pm 1 $ & 1.00 & 0.3750 \\
21 & 1.9 & $11 \pm 1 $ & 0.40 & 5.7125 \\
22 & 2.5 & $13 \pm 1 $ & 1.00 & 0.1191 \\
23 & 0.6 & $6 \pm 1 $ & 0.40 & 4.2900 \\
24 & 2.3 & $13 \pm 1 $ & 0.40 & 2.4630 \\
25 & 0.5 & $8 \pm 1 $ & 0.40 & 2.4293 \\
26 & 1.7 & $13 \pm 2 $ & 1.00 & 0.1094 \\
27 & 5.2 & $13 \pm 1 $ & 1.00 & 9.2550 \\
28 & 3.3 & $21 \pm 4 $ & 2.00 & 0.0214 \\
29 & 3.0 & $12 \pm 1 $ & 0.40 & 4.9275 \\
30 & 1.7 & $13 \pm 1 $ & 0.40 & 0.2722 \\
31 & 10.0 & $36 \pm 2 $ & 2.00 & 2.2778 \\
32 & 2.5 & $13 \pm 1 $ & 0.40 & 0.3292 \\
33 & 1.9 & $10 \pm 1 $ & 2.00 & 1.3217 \\
34 & 1.1 & $13 \pm 1 $ & 0.40 & 1.4917 \\
35 & 3.6 & $31 \pm 7 $ & 2.00 & 0.0322 \\
36 & 2.1 & $13 \pm 1 $ & 0.40 & 32.9000 \\
37 & 2.1 & $13 \pm 1 $ & 0.40 & 0.3735 \\
38 & 2.5 & $12 \pm 1 $ & 0.40 & 16.4300 \\
\\
\hline                                   
\\
A & 36.3 & $62 \pm 8 $ & 0.05 & 0.0003 \\
B & 6.3 & $7 \pm 1 $ & 1.00 & 33.1286 \\
C & 11.0 & $6 \pm 1 $ & 1.00 & 3.7817 \\
D & 158.0 & $333 \pm 70 $ & 0.40 & 1.2710 \\
E & 14.5 & $7 \pm 1 $ & 1.00 & 8.3714 \\
F & 39.8 & $199 \pm 98 $ & 0.40 & \num{1.40e-06} \\
G & 525.0 & $412 \pm 88 $ & 2.00 & 0.0679 \\
H & 7.6 & $7 \pm 1 $ & 1.00 & 65.9857 \\
I & 63.1 & $36 \pm 19 $ & 2.00 & 0.0035 \\
J & 275.0 & $506 \pm 351 $ & 0.05 & \num{2.61e-06} \\
K & 479.0 & $872 \pm 21 $ & 0.05 & 16.1000 \\
L & 398.0 & >371 & 0.05 & \num{1.29e-06} \\
M & -- & -- & -- & -- \\
N & 145.0 & >363 & 0.05 & \num{1.52e-06} \\
\hline                                   
\end{tabular}
}
\label{age_stellarmass_table}
\end{table}
%

\section{Discussion}\label{discussion}

\subsection{Ages and stellar masses of the young regions}

\begin{figure}
\centering
\includegraphics[width=\hsize]{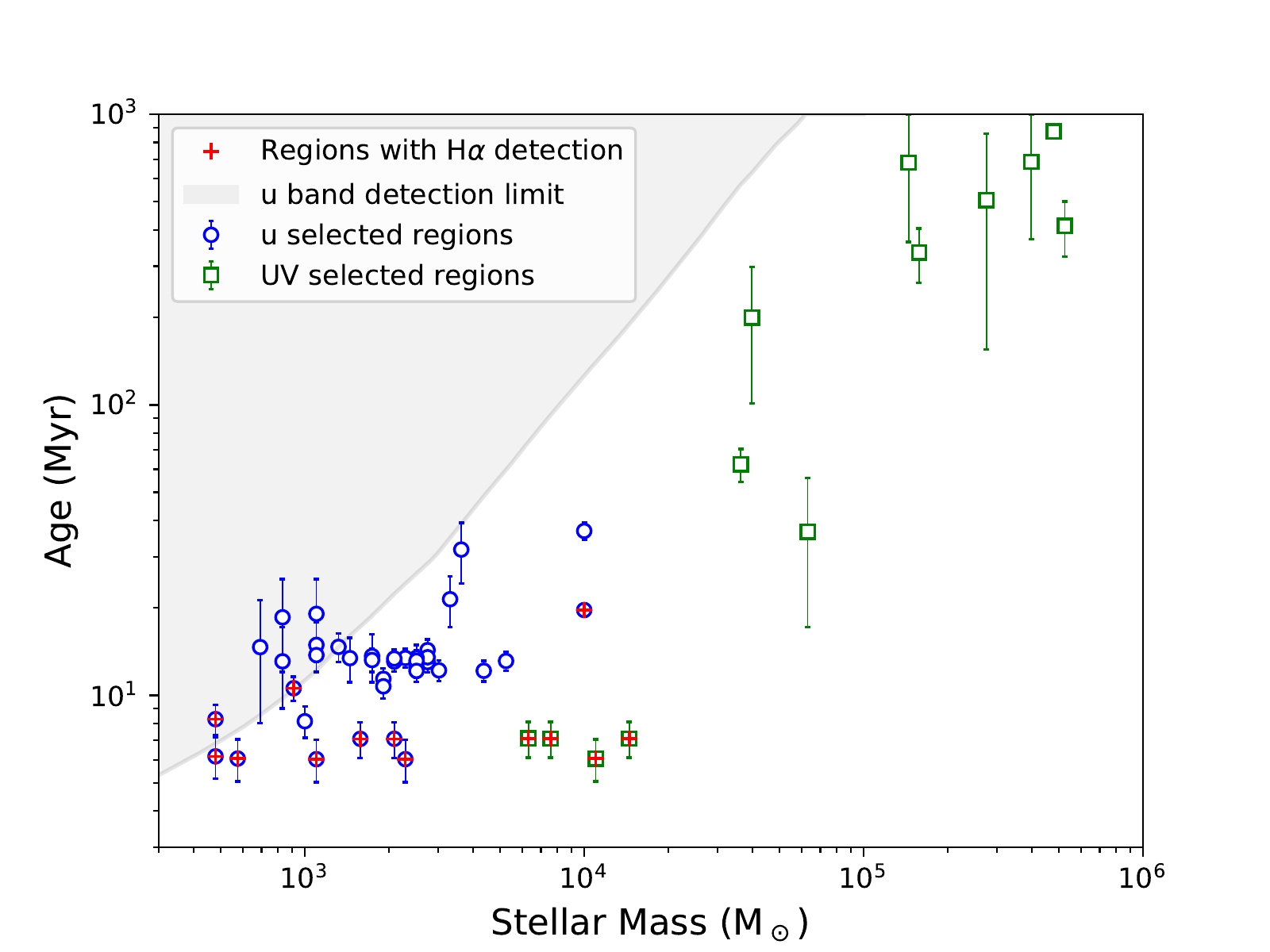}
  \caption{Age and stellar mass determined for all our selected regions. The blue circles and green squares represent \textit{u}-band- and UV-selected regions, respectively. The gray shaded area is our \textit{u}-band detection limit (lower mass or older clusters in the this area would not be detected based on the luminosity predicted by the Starburst99 models). The red crosses identify the \Ha{} detected regions.}
     \label{plot_age_vs_stellar_mass}
\end{figure}

The analysis performed in Sect. \ref{analysis}  shows that the majority of the star complexes associated with the HI gas cloud \agc{} located at a projected distance of $\sim$5 kpc from \udg{} have ages of a few tens of millions of years. These regions might therefore have formed within the gas removed from the \udg{} after a RPS event that started $\sim$200 Myr ago.
Figure \ref{plot_age_vs_stellar_mass} shows the estimated ages and stellar masses of all our selected regions. 
When the regions have an \Ha{} detection, the \textit{u}-selected and UV-selected regions have similar ages. In the absence of \Ha{} detection, the ages of the UV-selected regions tend to be larger than those of the $u$-selected regions, while UV emission is usually related to a younger population than $u$-band emission. However, the larger apertures of the UV regions make them more likely to be affected by any older underlying stellar population, and we reiterate the fact that we also expect some of them to be background sources.
For regions younger than 100 Myr (blue regions), the mean age of \textit{u}-band- and UV-selected regions are $14\pm1$ Myr and $21\pm4$ Myr, respectively (the uncertainty given in mean age is the formal error, rather than the dispersion of the age distribution). For a few of the fainter regions with low stellar mass and larger error bars, we are close to the \textit{u}-band detection limit on age, as can be seen in Fig. \ref{plot_age_vs_stellar_mass}. We obtain a total stellar mass of $9.1\times10^4$ \msun and $1.4\times10^5$ \msun for all the \textit{u}-band and UV-selected blue regions, respectively. The mean stellar masses of these blue regions are respectively $2.4\times10^3$ \msun and $2.3\times10^4$ \msun, which is within the mass range of $10^3-10^5$ \msun found in giant molecular clouds and HII regions of irregular galaxies \citep{kennicutt1989,fumagalli2011}. 
Such relatively low masses support the use of single generation populations to study them. Indeed, smooth extended star formation histories apply to systems including many molecular clouds and HII regions, while a single burst may better correspond to single clouds. Nevertheless, \mstar{}$ \approx 10^{4}$ \msun is close to the limit where the stochastic sampling of the IMF starts to play a role, resulting in over-estimation of ages using population synthesis models \citep{boselli18_vestige3}. However, our overall results would not be impacted if our ages were over-estimated because we already find many young regions.
It is clear from Fig. \ref{plot_age_vs_stellar_mass} that the UV-selected regions tend to have a higher stellar mass than the \textit{u}-band-selected regions. This can be attributed to their larger apertures which clearly capture a larger amount of the light emitted by star formation. 

Turning to similar studies in the Virgo cluster, 
\citet{fumagalli2011} constrained the ages along the blue tail of a dwarf irregular galaxy VCC 1217/IC 3418. Using optical and UV photometric bands, these latter authors performed SED fitting of the central galaxy and of the blue knots and filaments along its tail, assuming an extended star formation history. For the central galaxy, IC 3418, they found a star formation quenching time of $\sim$400 Myr due to RPS, but a large range of ages in the tail regions, from 80 to 1400 Myr. However, some of the star-forming regions were later spectroscopically confirmed as background objects by \citet{kenney2014}. 
For the confirmed tail regions, \citet{kenney2014} obtained ages ranging from 80 to 390 Myr, consistent with the quenching time from their models.
In the tail of NGC 4254, which is likely the result of tidal interactions, \citet{boselli18_vestige3} estimate an age of $\leq100$ Myr for typical star forming regions, whereas for the tail of a recently ram-pressure stripped ($\sim$50 Myr ago) dwarf galaxy IC 3476, \citet{boselli20_vestigeIX} give a typical age of $\leq20$ Myr for a few star forming complexes observed in the tail at $\sim$8 kpc from the stellar disk. 

Similar to these examples, the ages of the regions we obtained in our analysis
are young and point to recent formation (except for a few regions with very large ages and stellar masses that are likely contaminated by background objects). The ages of these young regions are consistent with the quenching of the disk occurring a few 100 Myr ago.

\subsection{Gradients along the tail}\label{section:gradients_along_the_tail}

\begin{figure}
\centering
\includegraphics[width=\hsize]{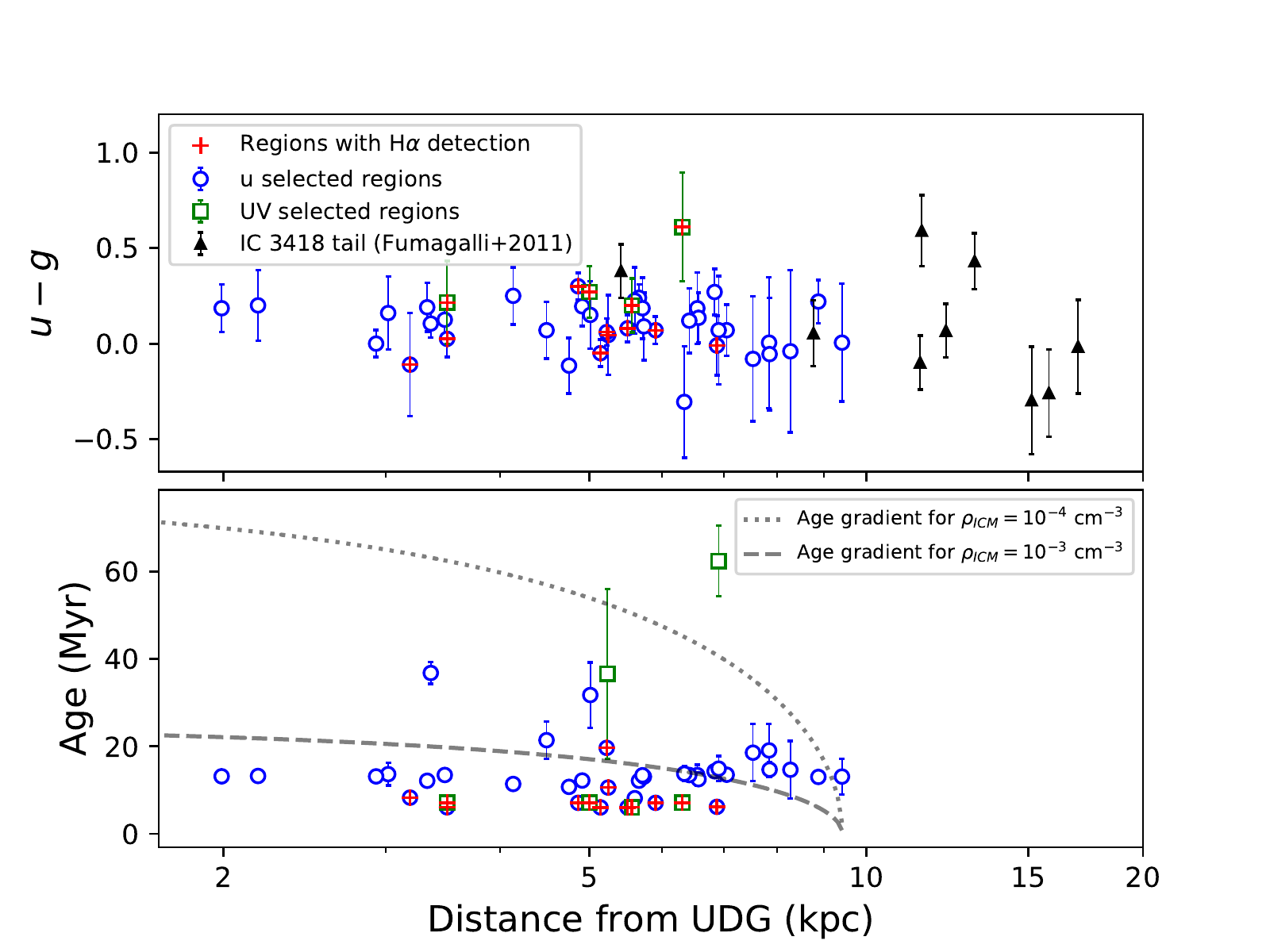}
  \caption{$u-g$ color and age of our blue regions as a function of their distance from the center of the UDG. Black triangles represent the knots and filaments from \citet{fumagalli2011}. The black dotted and dashed lines indicate the age gradients we measured for a 9.4 kpc stream following Eq. 2 of \citet{kenney2014}, for two different ICM densities with $\rho_{\mathrm{ICM}} = 10^{-4}$ and $10^{-3}$ cm$^{-3}$, respectively, as discussed in Sect. \ref{section:gradients_along_the_tail}.
  The blue open circles and green open squares mark our \textit{u}-band and UV-selected regions, respectively. The red crosses identify the \Ha{} detected regions.
}
     \label{plot_age_vs_distance}
\end{figure}

Figure \ref{plot_age_vs_distance} shows the $u-g$ color and age of the blue regions as a function of their projected distance from the center of the UDG. The top panel of Fig. \ref{plot_age_vs_distance} shows a comparison of the measured $u-g$ color of our regions with that of the knots of IC 3418 from \citet{fumagalli2011}. While our $u-g$ colors are consistent with theirs, we do not find any indication of a clear gradient, contrary to \citet{fumagalli2011} who observed a small color gradient in the tail of IC 3418, with the outermost part of the tail being relatively blue in comparison to the rest and at a larger radial separation than ours. %
With a large dispersion in the color and proximity to \udg{}, it is hard to draw strong conclusions as to the presence of a color gradient among our regions.

The bottom panel of Fig. \ref{plot_age_vs_distance} shows the age of the blue regions as a function of their distance from the center of the UDG. \citet{kenney2014} provided a relation for the age gradient from the head to the tail of a linear stream of fireballs (see their Eq. 2). We assumed a stream of length 9.4 kpc (the farthest region we observe), $\Sigma_{\mathrm{gas}} = 1 \text{\msun} \mathrm{pc}^{-2}$ in the outskirts of the UDG before undergoing RPS (obtained from the models discussed in Sect. \ref{analysis_rps}), and a relative velocity of $v = 1084$ km s$^{-1}$ for the HI gas cloud of \agc{} with respect to the Virgo cluster center \citep{boselli14_guvics,cannon2015}. We adopted two different values for the intra-cluster medium (ICM) density,  with  $\rho_{\mathrm{ICM}} = 10^{-4}$ and $10^{-3}$ cm$^{-3}$, corresponding to the ICM density at the distance of \udg{} from the cluster center \citep{Simionescu2017} and a typical ICM density of the Virgo cluster from \citet{vollmer2001}, respectively.
Using these values in Eq. 2 of \citet{kenney2014}, we obtain a gradient of a few tens of millions of years from the head to the tail of our stream.
The ages that we measure are consistent in order of magnitude with the expected age gradient for a stream of this length.
However, considering the uncertainties and the scatter of our data, it is difficult to determine an age gradient from the observations. Moreover it is not surprising to see a lack of age gradient because these regions are very young. Clear gradients are usually seen in galaxies that interacted slightly longer ago \citep{fumagalli2011}.
Also, while our data are consistent with some predictions, the uncertainties on the adopted parameters (gas density, ICM density, relative velocity) can lead to a wide range of possible gradients, as can be seen in Fig. \ref{plot_age_vs_distance} for two densities. 
Moreover, the \citet{kenney2014} formula may correspond to an ideal situation,
but the formation of star clusters is not necessarily 
a continuous function of the distance from the stripping event,
as found in the simulations of \citet{steyrleithner2020} in which 
star formation sets in not immediately after the stripping event but in the stream behind.

\begin{figure}
\centering
\includegraphics[width=\hsize]{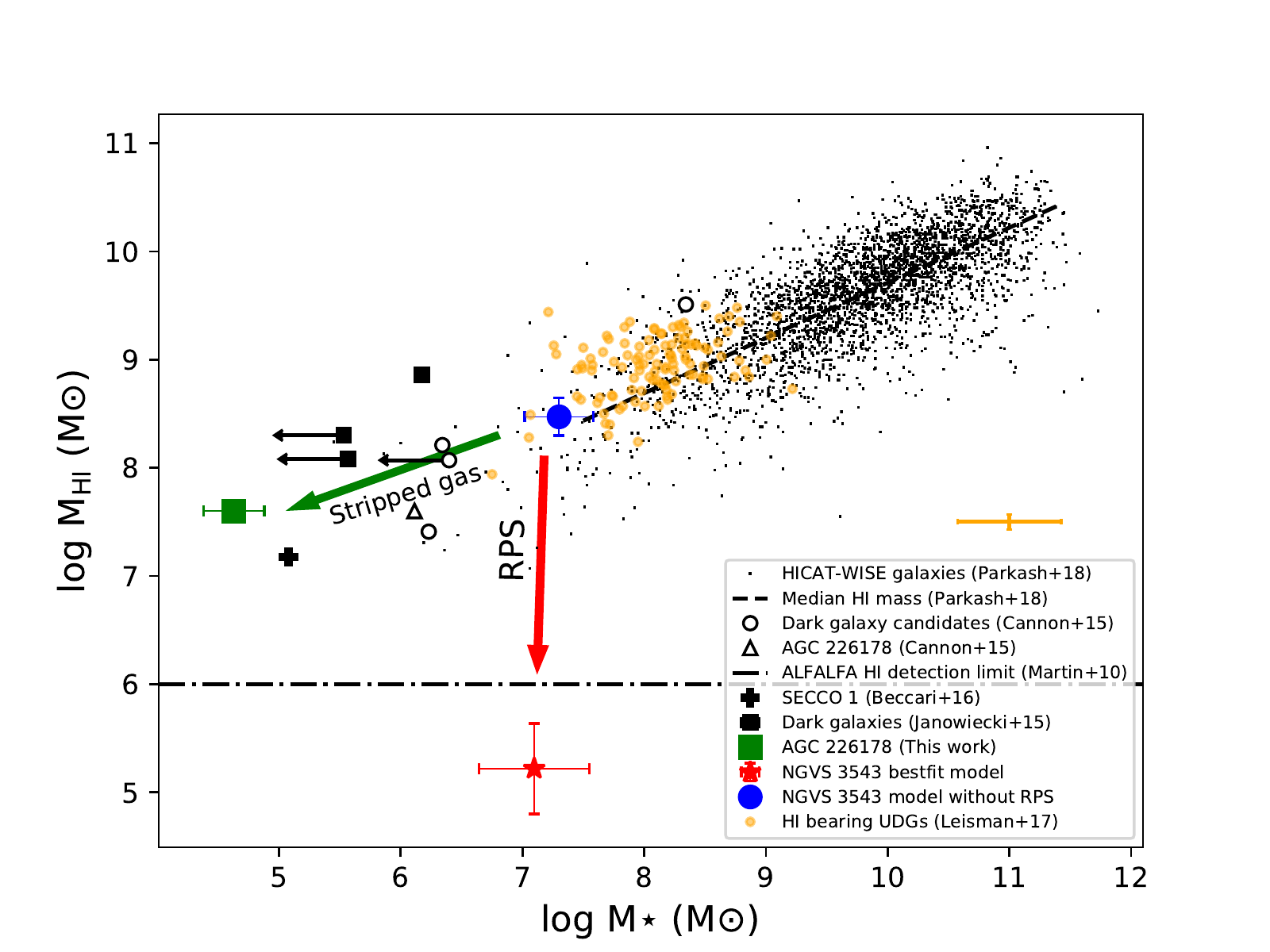}
  \caption{Stellar mass vs. HI mass of the HI-selected sample of spirals, UDGs and "dark" galaxies. The blue circle and the red star respectively mark the position of \udg{} before and after the RPS event, as given in Table \ref{galaxymodel_table}. The total gas masses from the models were converted into HI by multiplying by a factor 0.74 to take into account Helium and metals.
  Part of the stripped gas from the UDG can form \agc{}, shown as the green square.
  The small black squares show the dark galaxies from \citet{janowiecki2015}, the black open circles and triangle are those from \citet{cannon2015}. The black cross is SECCO 1 (AGC 226067) from \citet{beccari2017}. The black points and the black dashed line are the HI-selected spirals and their median HI mass, respectively, from \citet{parkash18}. The small yellow circles and error bars are the HI bearing UDGs and the mean dispersion from \citet{leisman2017}, respectively. The black dot-dashed line marks the HI mass detection limit of the ALFALFA survey \citep{ann_martin2010}.}
     \label{plot_mstar_vs_mhi}
\end{figure}

\subsection{A global scenario for the formation of almost dark galaxies and red UDGs}\label{global_scenario}

\subsubsection{An UDG undergoing a  RPS event}

The analysis of the ram-pressure models of \udg{} and the identification of some very young star forming regions near it lead us to an interesting question about the formation and evolution of such systems. 

Figure \ref{plot_mstar_vs_mhi} shows a comparison of the stellar mass vs. HI mass of our UDG and \agc{} with that of an HI-selected sample of regular spirals, UDGs and dark galaxies from the literature. 
\citet{leisman2017} provide a sample of HI-bearing UDGs from the ALFALFA survey. We estimated the stellar mass of these UDGs from their \textit{g}-band absolute magnitudes and $g-r$ color (see Table 1 of \citealt{leisman2017}), following the stellar mass-to-light-ratio--color relation for LSB galaxies given in \cite{du2020}. The distribution of UDGs falls along the low-stellar-mass tail of the \mstar{}-- $M_{\mathrm{HI}}$ relation for regular galaxies from \citet{parkash18}. These gas-rich, low-stellar-mass UDGs can be considered as the population of field blue UDGs discussed by \citet{prole2019}. The stellar mass and HI mass of our model  for \udg{} before the RPS event ($\sim$10$^7$\msun and $\sim$10$^8$\msun, respectively) suggest that its progenitor was similar to the population of blue UDGs.
The RPS event quickly transformed this galaxy %
into a gas-poor ($M_{HI} \sim 10^5$\msun), red UDG and totally quenched its star formation activity. Although our estimates are uncertain (as indicated by the error bars, not including additional sources of systematic error such as IMF choice), Figure \ref{plot_mstar_vs_mhi} illustrates this scenario in the global context of stellar and gaseous masses of galaxies covering a very large dynamical range.

The HI detection limit of the ALFALFA survey is of the order of $\sim$10$^{6}$\msun \citep{ann_martin2010}. The nondetection in HI for \udg{} \citep{cannon2015} is therefore  in complete agreement with our RPS scenario. 
Moreover, the HI mass of \agc{} from \citet{cannon2015} ($M_{HI} = 4 \times10^{7}$\msun) corresponds to $\sim$10\% of the mass of the gas expected to be stripped from \udg{} as indicated by the models. Considering that in similar RPS events a large fraction of the stripped gas can also change phase, becoming ionized gas before hot gas \citep{boselli16,boselli20_vestigeIX}, our analysis is consistent with the gas detected as \agc{} having been recently stripped from \udg{} during the RPS event.  

\subsubsection{Formation of an almost-dark object}

We made an estimate of the total stellar mass corresponding to the HI-source \agc{} using the combined stellar masses of our \textit{u}-band or UV-selected regions within the HI contour of \agc{} (as shown in Fig. \ref{color_image}). This gives an average stellar mass of $\sim$5$\times10^{4}$\msun for \agc{}. This stellar mass and HI mass is consistent with a sample of some other almost dark galaxy candidates from the literature \citep{cannon2015, janowiecki2015, beccari2017}. However, we see  in Fig. \ref{plot_mstar_vs_mhi}  that the stellar mass of \agc{} from \citet{cannon2015} is about 20 times larger than the value we obtained. \citet{cannon2015} and \citet{janowiecki2015} used standard mass-to-light-ratio--color relations to estimate their stellar masses. This naturally provides a more massive stellar mass than ours, which is not computed with a standard mass-to-light ratio, but is adapted to the young stellar population. More generally, one has to be cautious when considering the stellar masses derived for dark galaxies, whose distance and nature are not always certain. For example, in the case of AGC229385, \citet{janowiecki2015} give a stellar mass of $2\times10^{6}$\msun while \citet{brunker2019} provide $4\times10^{5}$\msun for the same object. 
This difference is due to the different distances they adopted, respectively 25 Mpc and 5 Mpc. In our case, the \Ha{} detection in VESTIGE provides a strong indication of cluster membership and distance of \agc{}.

While this suggests that knots of young stars may also be associated to other almost-dark galaxies if they formed in a similar way, this is not necessarily the case as star formation is not always present in RPS tails \citep{boselli16}, and once formed, the star complexes do not suffer RPS anymore and may decouple from the gas \citep{jachym2019}.

\subsubsection{The possibility of tidal interactions}

\citet{beccari2017} studied another interesting object in the Virgo cluster, SECCO 1, with similar stellar and HI properties to those encountered in \agc{}. 
SECCO 1 is characterized by similar compact regions dominated by a young stellar population, but does not have any evident nearby companion. \citet{bellazzini18} suggests a possible origin for SECCO 1 as a stripped gas cloud from an interacting triplet of dwarf galaxies $\sim$250 kpc away. In this scenario, the stripped gas cloud that formed SECCO 1 could have survived in the ICM for $\sim$1 Gyr before becoming an isolated object with ongoing star formation.

In our case, we do not find the presence of any massive interacting companion that could explain its properties (e.g., tidal interactions). We investigated the possibility of tidal interactions in \udg{} by looking for low-surface-brightness features extending beyond its effective radius. For the green dashed region shown in Fig. \ref{color_image} along the NE of \udg{}, we observed a faint network of high-frequency structures with a statistically significant detection (signal-to-noise ratio, $S/N = 8$) characterized by a \textit{g}-band surface brightness of $\mu_{g} = 27.6$ \magperarcsec{} \citep[the significance of such an estimate has been cross-checked with the photometric procedure described in][]{fossati18,longobardi20}. 
Nevertheless, the retrieved irregular structure could also be related to the fact that the UDG progenitor was a low-mass-star-forming system ---which are generally characterized by an irregular morphology--- and that a few 100 Myr was not sufficient for the stars to be redistributed into a smooth spheroidal distribution. We thereby consider tidal interaction negligible and conclude that RPS is the dominant process taking place in the galaxy.

\subsubsection{Summary}

Our analysis strongly suggests that \udg{} is in the process of transformation from a blue UDG into a red UDG by a RPS event. Because red UDGs are very frequent in nearby rich clusters \citep{koda15,munoz15,mihos15,roman_trujillo17,janssens17}, this suggests that RPS could be one of the major processes in the formation of gas-poor red UDGs. As in the case of \agc{}, our observations also suggest that RPS could be the mechanism responsible for the formation of the almost dark objects discussed in the literature \citep{duc2008,cannon2015,janowiecki2015,leisman2017,brunker2019}. Some gas-poor, faint, still undetected parent galaxies could exist in the vicinity of these almost dark objects (in the form of quenched UDGs similar to \udg{}, which was not detectable before data at the depth of those provided by NGVS became available).

\section{Conclusions}\label{conclusion}
We present a multi-wavelength study of the Virgo cluster ultra-diffuse galaxy \udg{} and its surroundings using optical, UV, and \Ha{} narrow-band imaging data from the NGVS, GUViCS, and VESTIGE surveys, respectively. We identified an over-density of blue compact regions located at $\sim$5 kpc south of the stellar disk of the galaxy, the majority of which were detected in \Ha{} and UV. These regions are embedded in a large ($\sim$10$^7$\msun) cloud of HI gas previously detected by ALFALFA and the VLA.
Our comparative analysis of the spectro-photometric properties of the UDG galaxy and of its associated extra-planar star-forming regions, combined with tuned multi-zone models of galaxy evolution, led us to the following conclusions:

\begin{itemize}
    \item  The UDG galaxy \udg{} has undergone a RPS event over the last few hundred million years, transforming it from a gas-rich, blue UDG to a gas-poor, red UDG. The predominance of red UDGs in clusters could be related to similar events at earlier times. %
    \item A fraction of the gas lost from the perturbed gas-rich UDG during the RPS event has undergone a star formation episode, forming compact young star clusters in the tail of stripped gas.
    These newly formed regions have a mean age and stellar mass of the order of 20 Myr and  $10^4$\msun, respectively, consistent with being byproducts of the recent RPS event.

    \item These young star complexes are located well inside an HI gas cloud of $\sim$10$^7$\msun, previously identified as an almost dark galaxy by \citet{cannon2015}.

\end{itemize}

While many mechanisms have been proposed in the literature for the formation of these peculiar families of objects populating nearby clusters (UDGs and almost dark clouds), our results indicate that RPS, already known to be a major process shaping galaxy evolution in young clusters, has recently had a major driving effect in the formation of the \udg{} system. This galaxy may be representative of other objects with similar characteristics, in which the same process has occurred, albeit in a more distant past.

Narrow-band \Ha{} imaging data gathered during the VESTIGE survey have been of paramount importance in the study of the star formation history of this peculiar system. We are therefore planning to extend this study of the origin of UDGs and LSB galaxies to the whole Virgo cluster once the survey is completed.

\begin{acknowledgements}
We are grateful to the whole CFHT team who assisted us in the preparation and in the execution of the observations and in the calibration and data reduction: 
Todd Burdullis, Daniel Devost, Bill Mahoney, Nadine Manset, Andreea Petric, Simon Prunet, Kanoa Withington.
We acknowledge financial support from "Programme National de Cosmologie and Galaxies" (PNCG) funded by CNRS/INSU-IN2P3-INP, CEA and CNES, France,
and from "Projet International de Coop\'eration Scientifique" (PICS) with Canada funded by the CNRS, France. Co-author Matteo Fossati has received funding from the European Research Council (ERC) under the European Union's Horizon 2020 research and innovation programme (grant agreement No 757535).
\end{acknowledgements}

\bibliographystyle{aa}
\bibliography{bibliography}

\onecolumn
\begin{appendix}
\section{\chisq{} distributions for the determination of stellar mass and age of the regions}\label{appendix:stellar_mass_age_chi2}
Figures \ref{appendix_fig_u_selected} and \ref{appendix_fig_UV_selected} give the \chisq{} fitting results used for the determination of stellar mass and age of all the \textit{u}-band- and UV-selected regions discussed in this work. See Sect. \ref{age_stellar_mass_determination} and Fig. \ref{chi2region_example} for a detailed description of the models and one example figure.

\begin{figure}[H]
    \centering
    \foreach \i in {1,...,15} {%
        \begin{subfigure}{0.3\textwidth}
            \includegraphics[width=\linewidth]{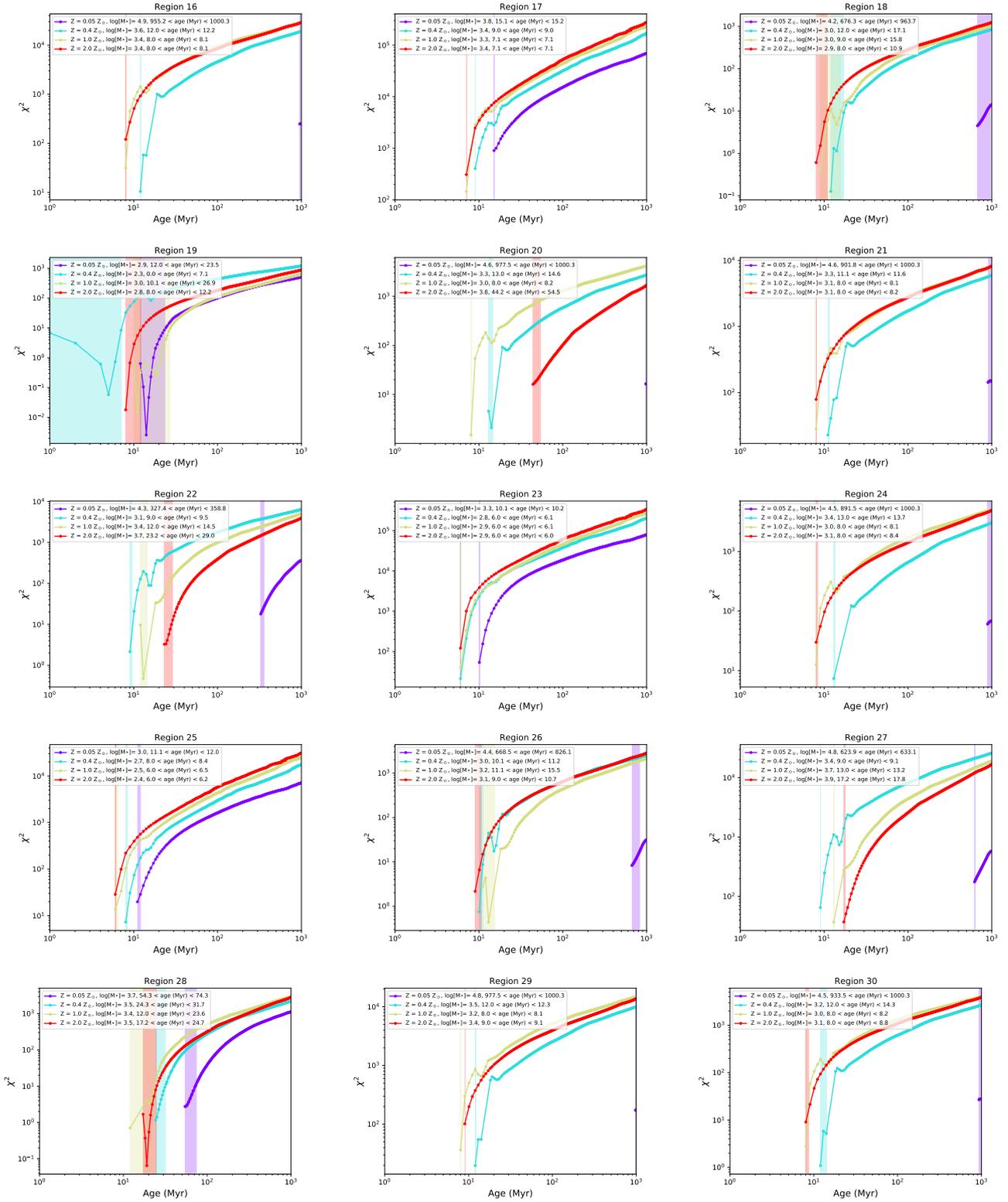}
        \end{subfigure}\quad}
\caption{\chisq{} as a function of age for the \textit{u}-band-selected regions. See Fig. \ref{chi2region_example} for an example and details in Sect. \ref{age_stellar_mass_determination}.}
\end{figure}

\begin{figure}[H]
    \centering
    \ContinuedFloat
    \foreach \i in {16,...,30} {%
        \begin{subfigure}{0.3\textwidth}
            \includegraphics[width=\linewidth]{chi2_images/u_selected/chi2_u_selected_region\i.pdf}
        \end{subfigure}\quad}
\caption{\chisq{} as a function of age for the \textit{u}-band-selected regions. See Fig. \ref{chi2region_example} for an example and details in Sect. \ref{age_stellar_mass_determination}.}\label{appendix_fig_u_selected}
\end{figure}

\begin{figure}[H]
    \centering
    \ContinuedFloat
    \foreach \i in {31,...,38} {%
        \begin{subfigure}{0.3\textwidth}
            \includegraphics[width=\linewidth]{chi2_images/u_selected/chi2_u_selected_region\i.pdf}
        \end{subfigure}\quad}
\caption{\chisq{} as a function of age for the \textit{u}-band-selected regions. See Fig. \ref{chi2region_example} for an example and details in Sect. \ref{age_stellar_mass_determination}.}
\end{figure}

\begin{figure}[H]
    \centering
    \foreach \i in {1,...,14} {%
        \begin{subfigure}{0.3\textwidth}
            \includegraphics[width=\linewidth]{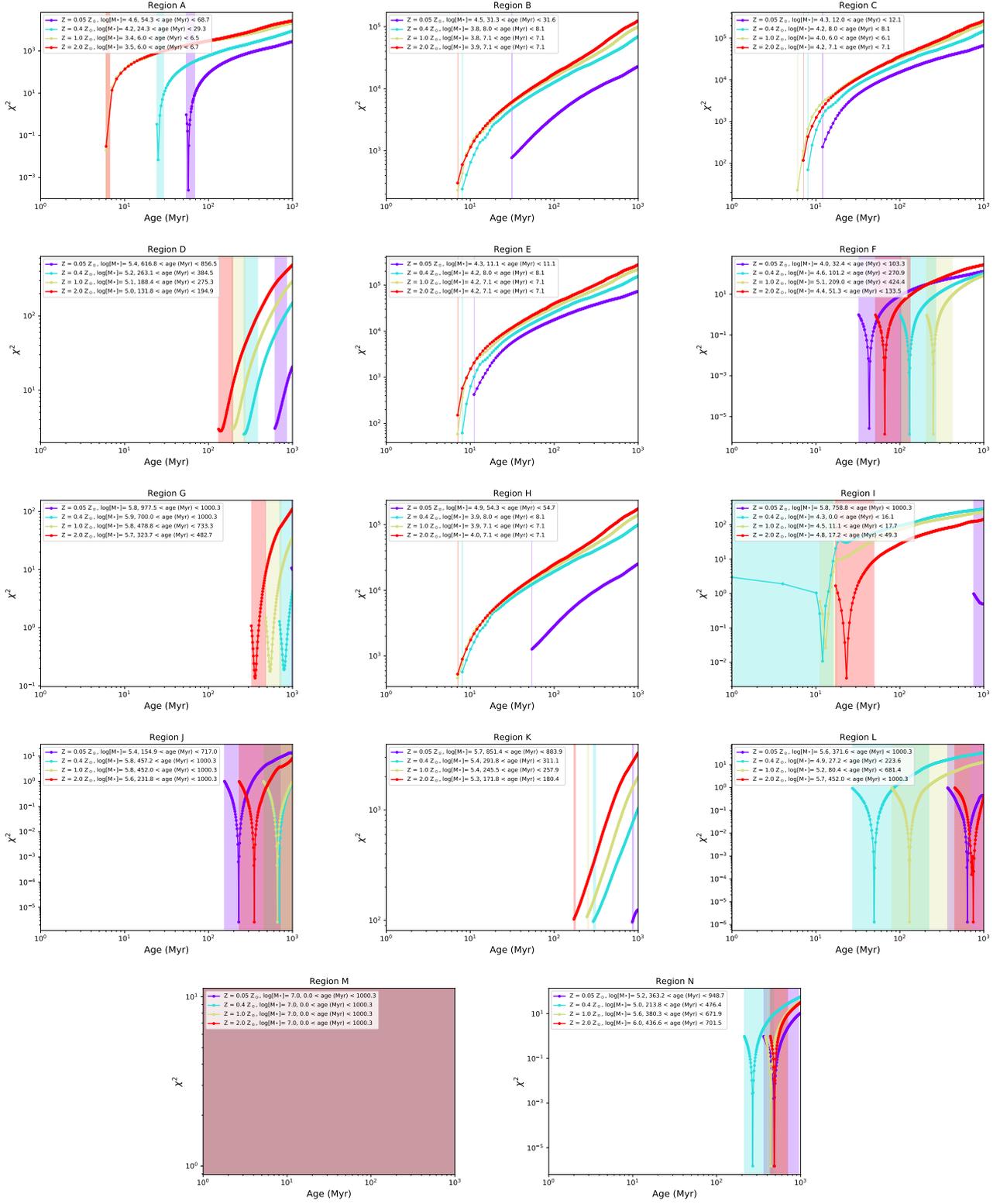}
        \end{subfigure}\quad}
\caption{\chisq{} as a function of age for the UV-selected regions. See Fig. \ref{chi2region_example} for an example and details in Sect. \ref{age_stellar_mass_determination}.}\label{appendix_fig_UV_selected}
\end{figure}

\newpage


\section{CIGALE determination of ages and stellar masses of \textit{u}-band- and UV-selected regions}\label{appendix:cigale_fitting}

Numerous assumptions in the stellar population models and fitting methodology could affect our results. To investigate this, we carried out a completely independent estimation of the properties of the regions using the SED modelling code CIGALE \citep{burgarella05,Noll09,boquien19}. We fitted each region given in Table \ref{measurements_table} with single burst population models from CIGALE, using the input parameters given in Table \ref{cigale_input} and following the same approach we adopted with Starburst99 models \citep{starburst99}. The differences with respect to the fit made with Starburst99 include the use of the \citet{chabrier03} instead of \citet{kroupa01} IMF, and different population synthesis models. Instead of relying on just the best-fit model, CIGALE estimates the physical properties from the probability distribution function. It also naturally takes into account upper limits in the computation of the goodness of fit. Finally, we performed two sets of fits, one with dust, and another without (as for Starburst 99).

\begin{figure*}[!hbt]
\centering
\begin{minipage}[b]{.46\textwidth}
\includegraphics[width=\hsize]{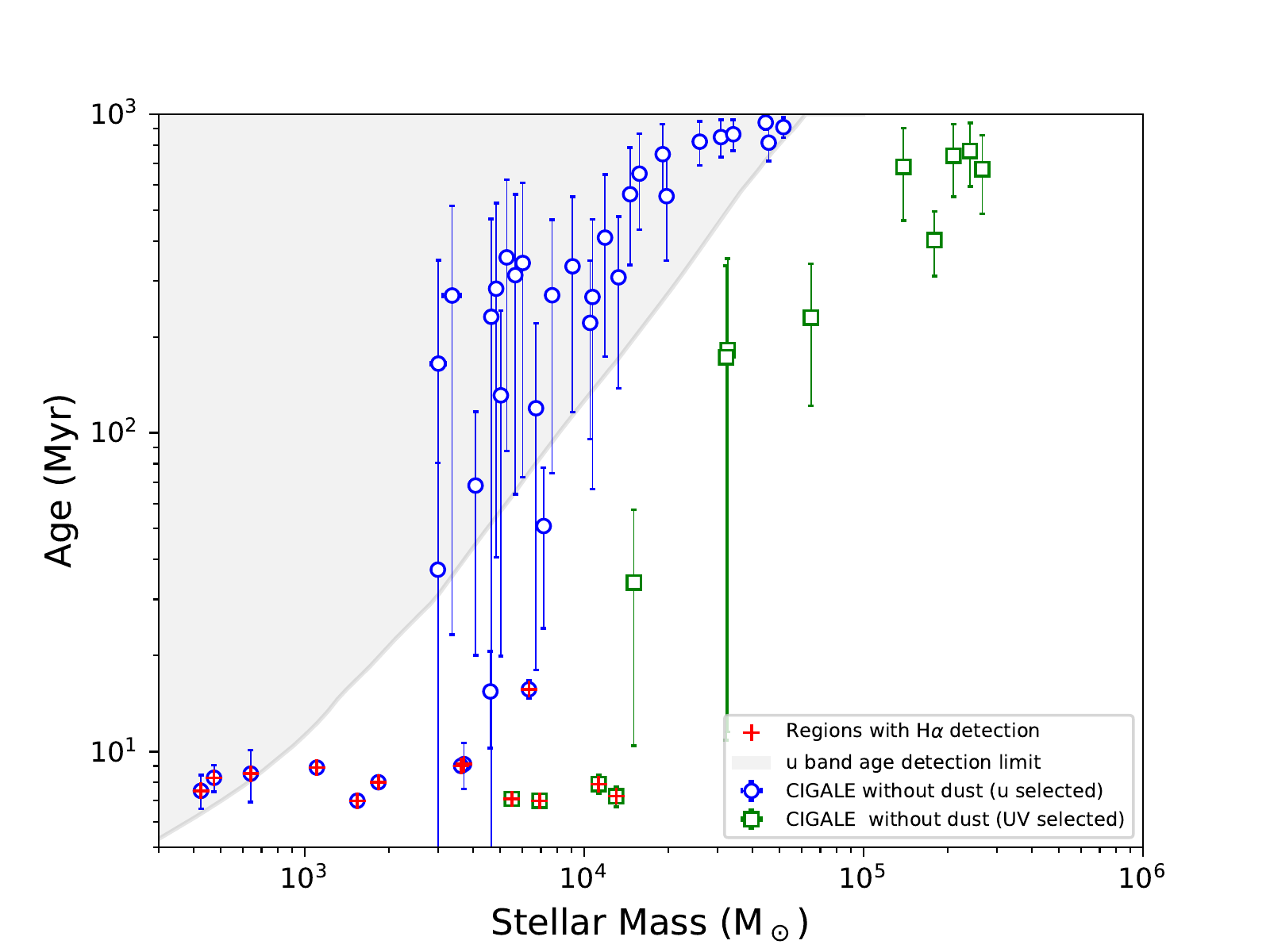}
\end{minipage}\qquad
\begin{minipage}[b]{.46\textwidth}
\includegraphics[width=\hsize]{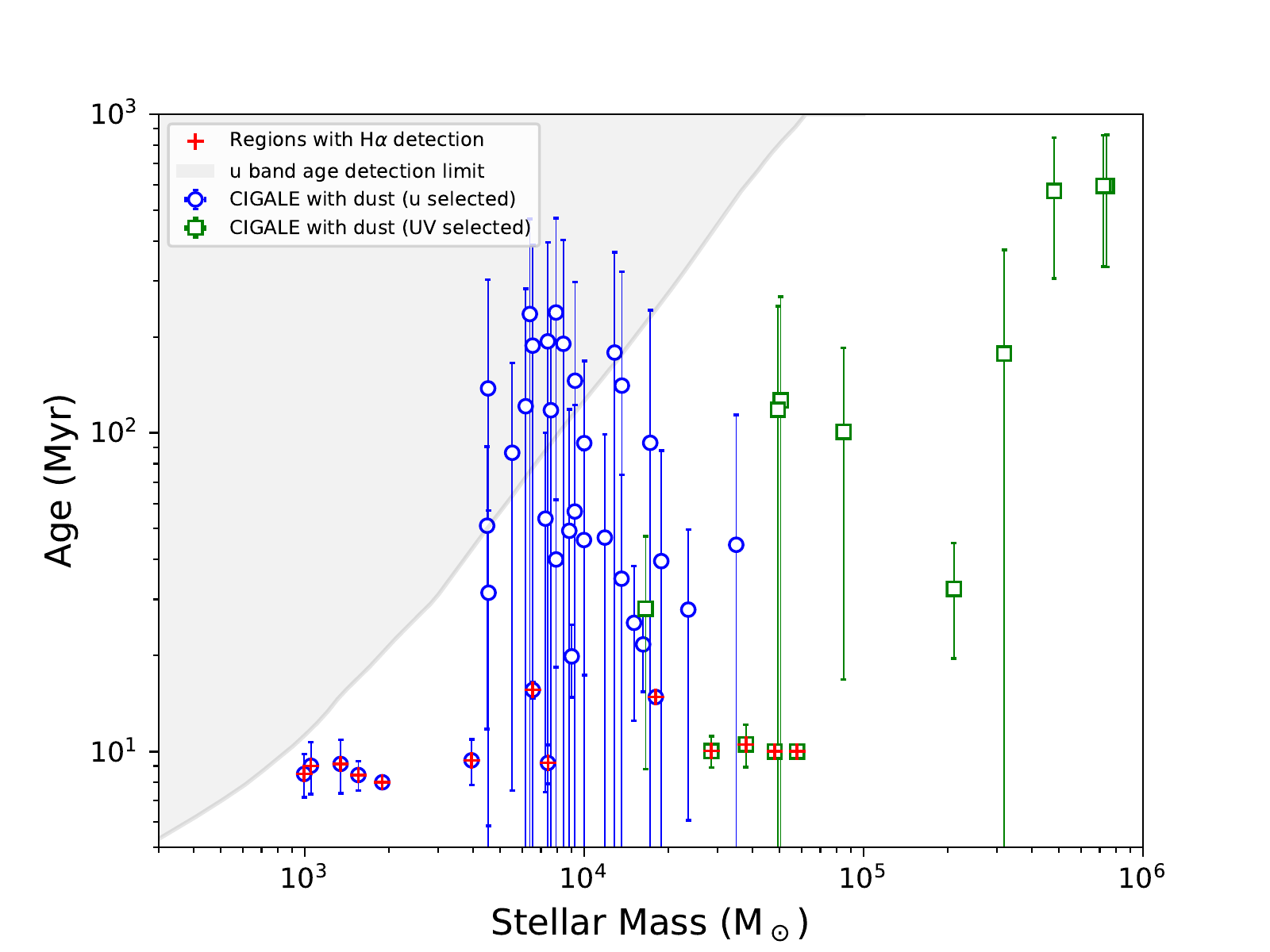}
\end{minipage}
\caption{Age and stellar mass determined using CIGALE for all the  regions in Table \ref{measurements_table}. These panels can be directly compared to Fig. \ref{plot_age_vs_stellar_mass} obtained with Starburst99. The left and right panels show the results obtained with CIGALE without and with dust, respectively. The blue circles and green squares represent \textit{u}-band- and UV-selected regions, respectively. The gray shaded area is our \textit{u}-band detection limit in stellar mass and age. The red crosses identify the \Ha{} detected regions.}
\label{appendix_plot:cigale_age_vs_mstar}
\end{figure*}

The stellar masses and ages obtained with CIGALE are shown in Fig. \ref{appendix_plot:cigale_age_vs_mstar}, and can be directly compared to those derived using Starburst99 (see Fig. \ref{plot_age_vs_stellar_mass}). CIGALE gives older ages and larger scatters than Starburst99 for many of the \textit{u}-band- and UV-selected regions. This effect can be due to the different spectrum of very young stellar populations between Starburst99 and \citet{bruzual_and_charlot03}. \citet{starburst99} stress that stellar evolutionary models are very uncertain when red super giant are important contributors, and the codes predictions may vary. It is especially the case in the age range 5-20 Myr for single star populations, and the reason for the presence of peaks in Fig. \ref{u-g_color_evolution_starburst99}, with a relatively red $u-g$ color around that age (these peaks are less large when \citet{bruzual_and_charlot03} populations are considered).
On the other hand, for regions with \Ha{} detection, CIGALE gives young ages, as we obtained with Starburst99. The \Ha{} measurements prove to be a very strong constraint in the modelling of very young star forming regions. 

Comparison between the two panels from Fig. \ref{appendix_plot:cigale_age_vs_mstar} shows that the inclusion of dust leads to   even younger ages.
Although our modelling using Starburst99 models did not account for dust, we obtained young ages ($<20$ Myr) for the majority of the regions. The inclusion of dust in these models would only produce even younger ages, like it is the case with CIGALE.

In conclusion, regardless of the code used for the stellar population, the fitting procedure, or the inclusion or not of dust,  we always find that a significant number of regions are indeed young  ($<20$ Myr), with similar stellar masses. We are therefore confident that our results are robust.

\begin{table}[!hbt]
\caption{Input parameters for CIGALE.}
\centering                          
\scalebox{0.93}{
\begin{tabular}{lll}
\hline \hline
& \multicolumn{1}{c}{\underline{\hspace{0.6cm}Model without dust\hspace{0.6cm}}} &
\multicolumn{1}{c}{\underline{\hspace{0.8cm}Model with dust\vphantom{g}\hspace{0.8cm}}}\\ 
Paremeter & & \\
\\ \hline
Pop. synth. mod.    & \cite{bruzual_and_charlot03}  & \cite{bruzual_and_charlot03} \\
Dust model          & No dust                       & \cite{calzetti2000} \\
IMF                 & \cite{chabrier03}             & \cite{chabrier03}     \\
Metallicity         & 0.004, 0.008, 0.02, 0.05      & 0.004, 0.008, 0.02, 0.05\\   
Age                 & 1 - 1000 Myr, step 1          & 1 - 1000 Myr, step 1\\
$E(B-V)$            & 0                             & 0 - 0.7 mag, step 0.01\\
UV bump amplitude   & 0                             & 0 \\
\hline
\noalign{\smallskip}
\end{tabular}}

\tablefoot{Line 1: Population synthesis model. Line 2: Dust model. Line 3: IMF. Line 4: Stellar metallicity. Line 5: Age interval and sampling. Line 6: Attenuation. Line 7: Amplitude of the UV bump \citep{Noll09}.}
\label{cigale_input}      
\end{table}

\end{appendix}

\end{document}